\documentclass[oneocolumn,showpacs,showkeys,preprintnumbers,amsmath,amssymb]{revtex4}

\usepackage{graphicx}
\usepackage{dcolumn}
\usepackage{bm}

\def\xder{{\dot x}}

\begin{document}


\title{ Relativistic quantum mechanics and \\
relativistic quantum statistics  based upon \\ a  novel perspective on relativistic transformation }

\author{Young-Sea Huang}
\email{yshuang@mail.scu.edu.tw}
 \affiliation{Department of Physics, Soochow University, Shih-Lin, Taipei 111, Taiwan
}


\begin{abstract}

   A novel perspective on relativistic transformation recently-proposed  provides an insight into the very meaning of the principle of relativity. With this novel perspective and Bell's theorem, we argue that  special relativity, instead of quantum theory,  should be  radically reformulated  to resolve  inconsistencies between those two theories. A new theory of relativistic quantum mechanics is  formulated upon this novel perspective. 
This new relativistic quantum mechanics is free from such anomalies as the negative probability density, the negative-energy states, Zitterbewegung,  and the Klein paradox deep-rooted in the current relativistic quantum mechanics. Moreover, a remarkable result is found that a particle  can not be confined within an infinite square well of width less than half of the Compton wavelength. As implications in nuclear physics, there is a lower bound on the size of atomic nucleus. Neither an electron, nor a positron, can be confined inside the nucleus by whatever interaction. 

  Furthermore, with this novel perspective, we argue that the postulates of non-relativistic quantum statistics fulfill the principle of relativity, as extended to the relativistic realm. A new theory of relativistic quantum statistics is formulated such that the probability distribution functions are the same as the well-known Maxwell-Boltzmann, Fermi-Dirac and Bose-Einstein distribution functions in non-relativistic quantum statistics. A relativistic speed distribution of a dilute gas is then derived by the new relativistic quantum statistics and the new relativistic  quantum  mechanics.  This relativistic speed distribution reduces to Maxwell speed distribution in the low temperature region. Yet, this relativistic speed distribution differs remarkably from J{\"u}ttner speed distribution in the high temperature region. Also, thermal properties of a dilute gas are studied by the new relativistic quantum statistics.

\end{abstract}

\pacs{
  Quantum mechanics, 03.65.-w; Special relativity, 03.30.+p;  Entanglement and quantum non-locality, 03.65.Ud;  Foundations of quantum mechanics, 03.65.Ta;  Philosophy of science, 01.70.+w, *43.10.Mq;   Nuclear models, 21.60.-n;  theory of quantized field, 03.70; 05.30-d quantum statistical mechanics; 05.20-y classical statistical mechanics. }

\keywords{conflict between  quantum theory and  special relativity, quantum theory, special relativity, the principle of relativity, Lorentz covariance,  non-locality, indeterminacy, Bell inequality, entanglement, realism, completeness, quantum mechanics, relativistic quantum theory,  Klein paradox,  {\it Zitterbewegung}, relativistic quantum field theory, Klein-Gordon theory,  Dirac theory, Atomic nucleus, relativistic quantum statistics, relativistic statistical physics, relativistic kinetic theory.}  

\maketitle

\section{\label{sec:1} Foundational conflicts between relativity theory and quantum theory}

Einstein's relativity theory and quantum theory are the two revolutionary theories evolving from classical physics. Nowadays,  they are  the two pillars of modern  physics,  for example,  relativistic quantum mechanics and  quantum electrodynamics  are constructed upon them.  In spite of  Einstein's relativity theory and quantum theory being,  respectively,  well verified  experimentally,  debates on the foundational conflicts between these two  theories seem  endless  \cite{ Albert,  Gisin2, Maxwell,Hilgevoord,Stefanovich, YSHuang, Seevinck,Maudlin, Sonego,Rem, Myrvold, Gisin, Ghirardi, Redhead,Jarrett, Nistico, Bahrami}. The perspectives on the space-time and the physical world adopted by classical physics, relativity theory, and quantum theory are briefly outlined in Table~\ref{tab0}.  Einstein's  relativity theory  overthrows Newtonian  absolute space and  absolute time as  presumed  in classical theory, whereas quantum theory is formulated  on that conventional notion of  space and  time.    On how the physical world  could  be described,   relativity theory adopts  some basic  notions of  classical theory:   there is no instantaneous action at a distance,  and  a physical theory  should be able to   predict definite values for all physical quantities pertaining to a system in the physical world.  Quantum theory,  in contrast, provides  some perspectives on the physical world  that are radically different from basic notions of classical theory:  non-locality,  quantum of action, indeterminacy, and the probabilistic interpretation of the wave functions. 
\begin{table}[ht]
\begin{tabular}{|c||l|l|} \hline 
   \text{ } & \hskip0.5cm \text{space-time concept} & \hskip1.5cm \text{view of physical world} \\ \hline\hline
 classical  physics & Newtonian space and time & determinism, realism and locality
 \\ \hline
 relativity  theory & Minkowski's space-time & determinism, realism and locality \\ \hline
 quantum  theory & Newtonian space and time & indeterminism, non-realism and non-locality \\ \hline
 \end{tabular}
\caption{\label{tab0}
Respective notions of the space-time and the physical world adopted by classical physics, relativity theory, and quantum theory are briefly outlined. Relativity theory and quantum theory are fundamentally incompatible due to their contradictory notions of the space-time and the physical world.  }
\end{table}

    Are relativity theory and quantum theory really incompatible?   Debates on this  subject are  related to  such issues  as  non-locality, realism and completeness,  raised by the well-known EPR article of  Einstein,  Podolsky and Rosen,  {\it  Can Quantum-Mechanical Description of Physical Reality Be Considered Complete?}  \cite{ Einstein}.   With the belief that  for each element of physical reality  a corresponding physical quantity should have a definite value prior to measurement (realism),  and their necessary condition for  completeness of a physical theory as well as criterion for an element of reality \cite{note1},  applying  the classical locality \cite{note2},  Einstein {\it et al }  showed that from the quantum-mechanical wave function of  two entangled particles, exact values of  either the position $Q$ or the momentum $P$ of  one particle  can be  inferred  from  respective  measurements on either the position $q$ or the momentum $p$ of the other particle   carried out at a  distantly separated place (far from the particles' interaction).  Then, by what they term  {\it reasonable definition of reality}, they  argued that a particle  possesses simultaneous elements of reality for the non-commuting observables of  position $Q$ and momentum $P$.  Yet,  according to Heisenberg's uncertainty principle, simultaneous values of  position and momentum of a particle can not be exactly determined by the wave function.  Supposing  quantum theory is correct,  either  quantum-mechanical description of physical phenomena  is not complete, or  these two non-commuting observables of  position  and momentum  can not have simultaneous elements of reality.  Thus, according to the EPR argument, there exist elements of physical reality whose corresponding physical quantities can not be predicted, with certainty, by quantum theory.  Einstein {\it et al } claimed that  quantum theory must at best be  incomplete.   An anticipated complete theory, with  additional variables not known yet,   should be able to predict simultaneous values of  position and momentum of a particle.    Bohr rebutted the EPR argument,  and  contended  "quantum-mechanical description of physical phenomena fulfills, within its scope,  all rational demands of  completeness"  \cite{Bohr}.

              To present the EPR argument in a comprehensible form,  Bohm and Aharonov  considered a special example --  a system of  two  spin-1/2 particles is created in the singlet  spin state and these two particles then fly off  in opposite    directions to two  detectors \cite{Bohm}.    Using  this special example,  Bell studied  the correlation of the spin states of the two particles.    Directions of spin measurement, on the entangled duo,  measured  by two spatially-separated detectors were  allowed to be oriented  arbitrarily,  instead of  along the same direction \cite{Bell}.  Bell  discovered that the correlation of  the  spins  of  the two entangled particles predicted by  any theory,  relativity theory included,  that requires the classical locality,  must obey   the so-called Bell inequality.   In contrast,  the  correlation of the spins  of the two particles  predicted by quantum theory violates the  Bell inequality.  Bell's discovery made those seemingly philosophical issues raised by the EPR argument  become testable by experiments.   Many experiments on  generalized Bell's  inequalities, for example, the CHSH inequality \cite{Clauser}, were  performed, and  their results  have so far been  in agreement with  the predictions of quantum theory  \cite{Aspect,Aspect1,Weihs,Tittel,Salart}.  Those  experimental tests  indicate  not merely that  relativity theory and quantum theory are incompatible,  but that  relativity theory is probably not valid.

       Either relativity theory and quantum theory  are incompatible  or  relativity theory is invalid  entail a   catastrophe  to the foundation of modern physics.  The validity of relativistic quantum theory and quantum electrodynamics  are called into question, since their  foundations  are insecure.    Many suggestions to resolve the conflict between quantum theory and relativity theory  have been proposed, but none of them are totally convincing so far.    It is still unclear how to get a consistent description of the physical world out of these two fundamentally incompatible theories \cite{Seevinck}.    Though most physicists seem unwilling to give up  relativity theory,  we think that  relativity theory should be  radically reformulated  to reconcile with quantum theory,  since experimental tests of  the Bell inequality so far indicate that relativity theory is probably wrong.

\section{\label{sec:2} A novel perspective on relativistic transformation}

       One major conflict between Einstein's relativity theory  and quantum theory is rooted in their mutually contradictory  notions of space and time \cite{Albert, Gisin2, Maxwell,Hilgevoord, Stefanovich, YSHuang, Seevinck,Maudlin,Rem,Sonego}.  According to Einstein's relativity theory,  simultaneity of two space-like events is no longer absolute,  that is,  the time order of the events depends on reference frames used to describe the events.  Contrarily, quantum theory is formulated  on Newtonian  absolute space and  absolute time.  Simultaneity is absolute in quantum theory;  without it,   Heisenberg's uncertainty principle and the probabilistic interpretation in quantum theory become meaningless. Moreover, experiments supporting Einstein's notion of space-time are not beyond question, and further experimental tests without controversies are needed \cite{Prohovnik, Kantor,Dingle,Essen,YSHuang6}.

 To be consistent with quantum theory,   an anticipated   new theory of relativity should  be necessarily formulated on the same  concept of space and time  as  is presumed in quantum theory.
A  new  relativistic transformation was   recently formulated on the  two postulates, the principle of relativity and the constancy of speed of light, the same as postulated in Einstein's special relativity \cite{YSHuang1}.

 Suppose that an inertial frame $X'$ moves with a velocity ${\bf V}$ along the positive $X^{1}$-axis with respect to another inertial frame $X$. At an instant of time a particle moving with a velocity ${\bf v}$ with respect to the frame $X$, the particle will move with a {\it virtual} spatial displacement $\delta {\bf x}={\bf v}\, \delta t$ during a  {\it virtual} infinitesimal time interval $\delta t>0$. We also define $\delta x^{0} \equiv c \,\delta t$, because the speed of light $c$ is an invariant. Thus, at that instant of time, the state of motion of the particle with respect to the frame $X$ can be characterized by  the infinitesimal four-displacement $\delta x^{\alpha}\, \equiv \, (\delta x^{0}, \delta {\bf x})$. This four-displacement $\delta x^{\alpha}$, the same as velocity ${\bf v}$, is defined at an instant of time. Moreover, this four-displacement is virtual, so it is different from $\Delta t = t _{2} - t _{1}$ and spatial displacement $\Delta {\bf x} \equiv {\bf x}(t_{2}) - {\bf x}(t_{1}) $ which is defined as difference of particle's real positions ${\bf x}(t_{2})$ and ${\bf x}(t_{1}) $ at two instants of time $t_2$ and $t_1$, respectively.   Similarly, the state of motion of a particle with respect to the frame $X'$ is characterized by  the infinitesimal four-displacement $\delta x'^{\alpha}\, \equiv \, (\delta x'^{0}, \delta {\bf x}')$.
By the postulates, the principle of relativity and the constancy of speed of light, the new  transformation  of relativistic four-displacements of a  particle between the two inertial frames $X$ and $X'$ is derived as 
\begin{equation}\label{eq1}
\left\{ \begin{array}{rcl}
          \delta x^{0}\!\!\! & = & \gamma (\delta x'^{0} + \beta \, \delta x'^{1}) \\
           \delta x^{1}\!\!\! & = & \gamma (\beta \, \delta x'^{0} + \delta x'^{1}) \\
          \delta x^{2}\!\!\! & = & \,\delta x'^{2}   \\
          \delta x^{3}\!\!\! & = & \, \delta x'^{3}\,\,\,  ,
         \end{array} \right. 
\end{equation}
where $\beta = V/c$  and $\gamma = 1/ \sqrt{1- \beta^2}$.
This new relativistic transformation of the virtual four-displacement $\delta  x^{\alpha}$ is named the differential Lorentz transformation (DLT). 

   The relativistic four-momentum of a free particle of rest mass $m$ moving with velocity ${\bf v}$ with respect to a frame is defined as $P^{\alpha} = m\,c\,  \delta x^{\alpha}/ \delta \tau $.  Here,  $ \delta \tau \equiv (\delta x^{\alpha}\, \delta x_{\alpha})^{1/2}=\delta x^{0}/ \gamma_{v}$ is an invariant under the differential Lorentz transformation, where $ \gamma_{v} \equiv 1/ \sqrt{1- (v/c)^2}$.  From $P^{\alpha}$  for the free particle,  we have the well-known relativistic energy $E=P^{0}\, c = \gamma_{v}\, mc^{2}$  and  relativistic momentum ${\bf P}=\gamma_{v}\, m\, {\bf v}$.
   Also, the  transformation of  relativistic four-momentum  between the inertial frames $X$ and $X'$ is    
\begin{equation}\label{eqMomentum}
\left\{ \begin{array}{rcl}
          P^{0}\!\!\! & = & \gamma (P'^{0} + \beta \, P'^{1}) \\
           P^{1}\!\!\! & = & \gamma (\beta \, P'^{0} + P'^{1}) \\
          P^{2}\!\!\! & = & \,P'^{2}   \\
          P^{3}\!\!\! & = & \, P'^{3}\,\,\,  .
         \end{array} \right. 
\end{equation}
Equivalently,  the DLT can be considered as a transformation in the  space of relativistic four-momentum.

 According to Einstein's special relativity,   the currently-accepted relativistic transformation is the  Lorentz transformation (LT) of space-time coordinates   
\begin{equation}\label{eq2}
\left\{ \begin{array}{rcl}
          x^{0}\!\!\! & = & \gamma (x'^{0} + \beta \, x'^{1}) \\
           x^{1}\!\!\! & = & \gamma (\beta \, x'^{0} + x'^{1}) \\
          x^{2}\!\!\! & = & \,x'^{2}   \\
          x^{3}\!\!\! & = & \, x'^{3}\,\,\, .
         \end{array} \right. 
\end{equation}
The  DLT of displacements  Eq.~(\ref {eq1}) is usually thought as just a derivative of  the LT of space-time coordinates Eq.~(\ref {eq2}). Yet according to the novel perspective on relativistic transformation, the infinitesimal quantities $\delta x^{\alpha}$ in the  DLT are not the differential of the space-time coordinates $x^{\alpha}$ in the  LT. The DLT is compatible with Heisenberg's uncertainty principle, whereas the LT is not. Moreover, it should be emphasized that the new  relativistic transformation presumes  Newtonian  absolute space and  absolute time, whereas Einstein's special relativity does not. The DLT and the  LT are not equivalent, contrary to the current perspective.

 To  explicitly illustrate that  the  DLT and the LT are not equivalent, we pointed out an anomaly as induced by  the LT --- the problem of negative frequency of waves \cite{YSHuang2}.  Consider light waves propagating  in a medium which moves at "superluminal" speeds  opposite to the propagation direction of  waves. 
 \begin{figure}[h]
\begin{center}
\includegraphics[width=0.45\textwidth, clip=]{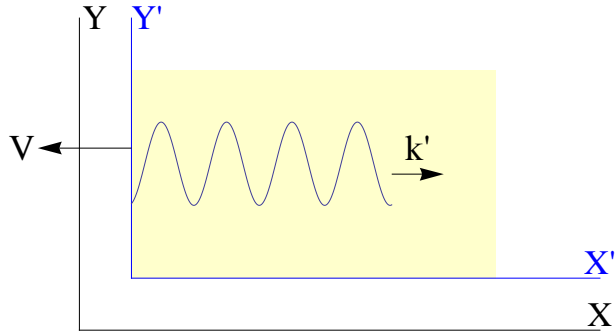}
\end{center}
\caption{\label{fig1}
Light waves propagate with the wave vector ${\bf k}'$  in the positive x-axial direction, relative to the medium rest frame $X'$. The  frame $X'$ moves with velocity ${\bf V}$ in the negative x-axial direction with respect to the frame $X$.  }
\end{figure}
      Referring to Fig.~\ref{fig1},  relative to the medium rest frame $X'$,  light waves propagate in the positive x-axial direction with speed $v'=\omega'/ k'$, where $\omega'$ is the frequency of the waves ($\omega'>0$),  and $k' $ is the wave vector ($k' >0$).  The frame $X'$ moves with velocity $V$ in the negative x-axial direction with respect to the frame $X$, and $V> v'$.  Then, by the LT, together with the assumption of  invariance of the phase of waves,  $\omega\, t -{\bf k}\cdot {\bf r} = \omega'\, t' - {\bf k}'\cdot {\bf r}'$,   we have  $\omega = \gamma\, ( 1- V/v') \, \omega' <0$ and $ k = \gamma \, (1-Vv'/c^2)\, k' >0$,  for the light waves propagating with respect to the frame $X$ . That is, relative to the frame $X$,  the light waves propagate in the positive x-axial direction, but with  {\it negative} frequency.  However,  light waves can not oscillate with negative frequencies. This anomaly is resolved by the  new relativistic transformation  \cite{YSHuang2}.

 Furthermore, we pointed out that the current transformation of Maxwell's equations of electrodynamics  by the LT does not truly fulfill the principle of relativity \cite{YSHuang3}. The reason is as follows: The original electromagnetic fields ${\bf E}({\bf r},t)$ and ${\bf B}({\bf r},t)$ relative to one inertial frame $X$ are  defined at the same time; yet by the LT, the transformed electromagnetic fields ${\bf E}'({\bf r}',t')$ and ${\bf B}'({\bf r}',t')$ relative to another inertial frame $X'$ are not defined simultaneously, since the LT is a transformation of space-time coordinate  \cite{Jackson}. That is, the time $t'$ in the transformed fields ${\bf E}'({\bf r}',t')$ and ${\bf B}'({\bf r}',t')$ are not at the same time. This non-simultaneity due to the  LT poses serious problems of  interpreting  those transformed fields defined over time extending from past to  future.

 The requirement of Lorentz covariance of the laws of physics is not sufficient for the laws to fulfill the principle of relativity \cite{Oyvind}. Besides such covariance requirement, all quantities in the covariant equations must be defined and interpreted physically in the same way in all inertial frames. The transformed fields ${\bf E}'({\bf r}',t')$ and ${\bf B}'({\bf r}',t')$ relative to the frame $X'$ are not  defined and interpreted physically in the same way as the original fields ${\bf E}({\bf r},t)$ and ${\bf B}({\bf r},t)$ relative to the frame $X$. The  LT renders Maxwell's equations Lorentz covariant only in a  superficially consistent  manner. The usual way of rendering Maxwell's  equations  form-invariant by the LT does not truly fulfill the principle of relativity.

\begin{figure*}[h]
\begin{center}
\includegraphics[width=0.8\textwidth, clip=]{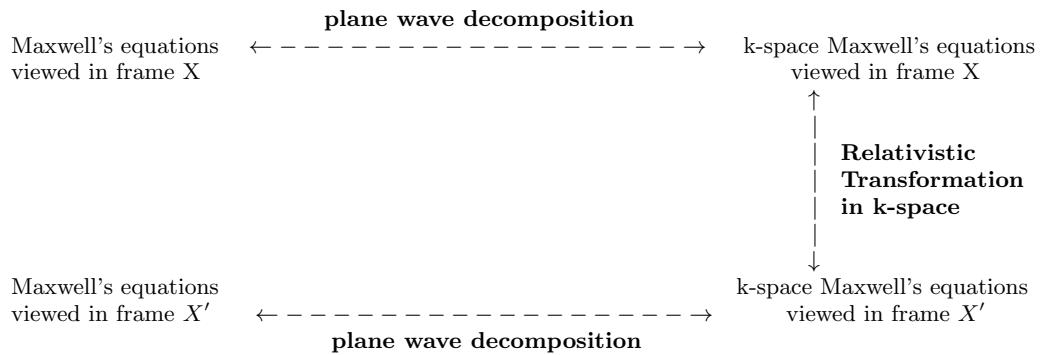}
\end{center}
\caption{\label{fig2}
Maxwell's equations are made form-invariant by the new scheme of relativistic transformation, i.e.,  transformation in the ${\bf k}$-space, rather than  the space-time space.}
\end{figure*}

  In contrast, with the novel perspective on relativistic transformation,  Maxwell's equations were shown  form-invariant among inertial frames,  via transformation in the k-space,  rather than the space-time space \cite{YSHuang3}. Referring to Fig.~\ref{fig2}, first  by plane wave decomposition, from Maxwell's equations in the space-time space, we have corresponding Maxwell's equations in the k-space. The forms of Maxwell's equations in the k-space remain the same in all inertial frames, since the same Maxwell's equations hold in all inertial frames. Secondly, by transforming  electromagnetic fields in the k-space and then reversing  plane wave decomposition to construct the transformed electromagnetic fields in the space-time space, Maxwell's equations are shown form-invariant. Since this scheme of transformation is carried out at an instant of time, the transformed  electromagnetic  fields in the space-time space are defined at the same time. Thus, the transformed electromagnetic fields  are  defined and interpreted physically in the same way as the original electromagnetic fields.  The new relativistic transformation of Maxwell's equations truly  fulfills  the principle of relativity.  

\section{\label{sec:3} The principle of relativity and quantum theory}

 Can quantum theory be consistent with the principle of relativity?   According to the current notion of the principle of relativity  in Einstein's relativity theory,  mathematical formulas  of physical laws must be  Lorentz covariant  under the LT of  space-time coordinates.  Then,  physical laws  are required to be expressed as mathematical formulas of  space and time coordinates in order  to see whether or not  their mathematical formulas   satisfy this  Lorentz covariance criterion of the principle of relativity.  Suppose that a  physical law can not be expressed in terms of the space and time description,  then by what criterion  are  we able to determine whether or not this law fulfills the principle of relativity?  For instance,  the law of conservation of energy-momentum is not expressed in terms of the space and time description.  Yet, without a doubt,  this law fulfills the principle of relativity. Ultimately, the constancy of speed of light (one of the two postulates of relativity theory) fulfills the principle of relativity, despite not being LT expressed in terms of the space and time description.  
  
      According to quantum theory,  a physical system  is described by  a state in Hilbert space subject to certain laws.   For instance, in quantum theory  the spin of an elementary particle  is unlike a spinning top in  classical theory which has three well-defined  components in the three dimensional physical world.  Instead, the spin of  an elementary particle  is described as a state $| {\bf s} \rangle$ in Hilbert space obeying the spin rule $\hat{\bf s} \times \hat{\bf s} = i\, \hbar\, \hat{\bf s}$, where $\hbar = h/ 2 \pi$, and $h$ is Planck's constant.  In addition,  a system of identical particles is described by a state in a multi-dimensional configuration space  obeying Pauli exclusion principle and quantum statistics of identical particles.   Quantum statistics,  the spin rule,  Pauli exclusion principle and Heisenberg's uncertainty principle  are  not  expressed  as  functions  of   space and time coordinates.  Therefore,  by the Lorentz covariance  criterion,  it is impossible to see whether or not these  laws of quantum theory fulfill the principle of relativity.  With the current  interpretation of the principle of relativity, it is impossible to answer the question -- Can quantum theory be consistent with the principle of relativity? 
 
 Are  Planck's constant, Heisenberg's uncertainty principle, the spin rule,  Pauli exclusion principle and quantum statistics the same in all inertial frames?    Physical laws of  a true  theory must be the same in all inertial frames.   Suppose  quantum theory is indeed true for the description of physical phenomena, then laws of quantum theory must  fulfill the principle of relativity.   Consequently,  the current notion of  the principle of relativity  must be radically rectified.   
   A simple example is employed to illustrate that Heisenberg's uncertainty principle fulfills the principle of relativity via the novel perspective on relativistic transformation.  Consider a free particle moving with a definite momentum ${\bf p}$ and energy $E$ with respect to a frame $X$.   In quantum theory the motion of the particle is described as a state $| {\bf p} \rangle$ in Hilbert space.  
By Heisenberg's uncertainty principle $[ \hat x_{j}, \hat p_{k}]=i\, \hbar \, \delta_{jk}$,  the position observable $\hat x_{j}$  does not commute with the momentum observable $\hat p_{j}$.  Thus,  the exact position of the particle can not be exactly determined without disturbing the original state  $| {\bf p} \rangle$.  Suppose that this state is described in terms of   an abstract wave function  $\psi({\bf r},t) = \langle {\bf r}  | {\bf p}  \rangle$   with respect to the frame $X$.  The position of the particle with respect to the frame $X$ can only be predicted probabilistically as proportional to $|\psi({\bf r},t)|^2$.   With respect to another frame $X'$ uniformly moving relative to the frame $X$, the particle moves with a definite  momentum ${\bf p}' $ and energy $E'$,  by the  transformation of relativistic four-momentum,  instead of space-time coordinates.  With respect to the frame $X'$, the particle is in the quantum state $| {\bf p} ' \rangle$. Then, the wave function of the particle is $\psi'({\bf r'},t') = \langle {\bf r'}  | {\bf p} ' \rangle$ with respect to the frame $X'$. Also, the position of the particle with respect to the frame $X'$ can only be determined  probabilistically as proportional to $|\psi'({\bf r'},t')|^2$.    It should be emphasized that 
the wave function $\psi'({\bf r'},t')$ of the particle with respect to the frame $X'$ is not directly obtained from the wave function $\psi({\bf r},t)$ with respect to the frame $X$ by the  LT of  space-time coordinates.  Rather, the wave function  of the particle is transformed via the transformation of relativistic four-momentum $P^{\alpha}$. Heisenberg's uncertainty principle is satisfied with the novel perspective on relativistic transformation. Heisenberg's uncertainty principle holds in all inertial frames,  though its mathematical formula is not manifestly Lorentz-covariant.
 
  Therefore,  to fulfill the principle of relativity,  physical laws are required to be the same in all inertial frames,  rather than their mathematical formulas  are Lorentz-covariant under the LT of space-time coordinates.    To render quantum theory consistent with the principle of relativity,   laws of quantum theory are {\it ab initio} hypothesized to be the same in all inertial frames,  though their mathematical formulas  are  not  manifestly Lorentz-covariant in the usual sense.  Yet,  the validity of this hypothesis depends on the results of  experiments carried out on  consequences derived from this  hypothesis.

 \section{\label{sec:4} The principle of relativity and the Lorentz covariance criterion}

    According to the principle of relativity,  all inertial frames are equivalent, and thus the same physical laws hold in all inertial frames \cite{Giannetto,Scribner,Einstein2}.   However, there are different  viewpoints on how physical laws should be formulated in order to fulfill the principle of relativity \cite{Houtappel, Arunasalam,Phipps,Szabo}.     The most accepted viewpoint  is the Lorentz covariance criterion --  to fulfill the principle of relativity,  the mathematical formula of a physical law must be  Lorentz covariant  under the LT of  space-time coordinates \cite{Einstein2,Feynman,Norton}.    Nonetheless, as mentioned previously in Sec. \ref{sec:2},  the  Lorentz covariance criterion is not a sufficient condition for a physical law to fulfill the principle of relativity.   A physical law whose mathematical formula apparently satisfies the Lorentz covariance criterion  does not guarantee  that law fulfills the principle of relativity.  
       
    Furthermore,   it was pointed out that the manifestly covariant equation $\partial _{\alpha}  A^{\alpha} =0$ does not imply $A^{\alpha}$ is a Lorentz-covariant 4-vector.  It is possible that the same equation  $\partial _{\alpha}  A^{\alpha} =0$  holds  in all inertial frames,  but that equation is not Lorentz covariant subject to the Lorentz covariance criterion,  as in the case that the quantity $A^{\alpha}$ in that equation is not covariant  under the LT of  space-time coordinates \cite{YSHuang5}. That is, a physical law  may fulfill the principle of relativity,  but its mathematical formula  does not satisfy the  Lorentz covariance criterion. Therefore, the Lorentz covariance criterion  is not a necessary condition for a physical law to fulfill the principle of relativity.   
    
    A strict and universal application of  the  technique of covariance  may result in  mathematical formalisms that have no physical significance whatsoever \cite{Wolfgang,Kretchmann}. Even though mathematical covariance in principle imposes no physical significance, the Lorentz covariance criterion is still applied to  formulate  relativistic   physical laws \cite{Schwarz, Rowe}.  Now, based on the novel perspective on relativistic transformation, physical laws, for example, Maxwell's equations, are rendered form-invariant via   transformation  of physical quantities,  instead of  space-time coordinates. The Lorentz covariance criterion is not  essential  to  formulate   relativistic   physical laws.

\section{\label{sec:5} Lagrangian relativistic mechanics based on the novel perspective on relativistic transformation }

  With the novel perspective on relativistic transformation, quantum theory and relativity theory can be integrated harmoniously. Next, we will formulate a new theory of relativistic quantum mechanics, and apply it to study standard  problems in quantum mechanics, square step and barrier potential. 

  A new relativity theory is formulated by the novel perspective on relativistic transformation, and the equation of motion  is 
\begin{equation}\label{rceq}
m\, \frac{ d^{2}{\bf r}}{ dt^{2}}={\bf F}\,\,(1-({v \over c})^{2}),
\end{equation}
where $m$ is the mass of the particle, ${\bf r}$ its position, $v$ its speed and  $c$ the speed of light \cite{yshuang2}. 
This equation of  motion has co-directional force and acceleration. Also, it shows that  acceleration of a particle varies with its speed, decreasing as its speed approaches the speed of light. Particles can not be accelerated to a speed over the speed of light.  This equation of motion contains Newton's equation of motion
\begin{equation}\label{ceq}
m\, \frac{ d^{2}{\bf r}}{ dt^{2}}={\bf F},
\end{equation}
as a low-speed limit. 
Yet, Eq.~(\ref{rceq}) is different from the equation of motion  in Einstein's special relativity 
\begin{equation}\label{Einreq}
\frac{d \, (\gamma\, m\, {\bf v})}{ dt}={\bf F},
\end{equation}
where $\gamma = 1/ \sqrt{1-({v/ c})^{2}}$ . In contrast to Eq.~(\ref{rceq}) and Eq.~(\ref{ceq}),
according to Eq.~(\ref{Einreq}), acceleration and force are in general not along the same direction  \cite{Moller}.
Experiments are proposed to convincingly distinguish between these two equations Eq.~(\ref{rceq}) and Eq.~(\ref{Einreq}), since there is no experimental evidence that  differentiates the minute difference between the two equations \cite{yshuang3}.

 Suppose a particle moving in a conservative field  ${\bf F}({\bf r})= - \nabla V({\bf r})$,  where  $V({\bf r})$ is potential energy. From Eq.~(\ref{rceq}), we have
\begin{equation}\label{rceq2}
\int  {m \over 1-({v \over c})^{2}} {d^{2}{\bf r} \over dt^{2}}\cdot d{\bf r} = \int - \nabla V({\bf r})\cdot d{\bf r}.
\end{equation}
 After integration, we obtain
\begin{equation}\label{energyE}
 E = m\, c^{2}\,\, {\it ln}\, \gamma +  V({\bf r}).
\end{equation}
Here, the integration constant  $E$ is considered as the total energy of the particle including the kinetic energy  $m\, c^{2}\,\, {\it ln}\, \gamma$  and the potential energy $ V({\bf r})$. The kinetic energy is approximately equal to $ m \, v^{2}/2$, as $ v << c$. It should be noted that the total energy $E$ does not contain the rest mass energy $m\, c^2$, as $m\, c^2\, ln 1 =0$.

    From Eqs.~(\ref{rceq}) and (\ref{energyE}), we have
\begin{equation}\label{rceq3} 
m\, \frac{d^{2}{\bf r}} { dt^{2}}= - e^{-2(E-V({\bf r}))/m c^{2}} \nabla V({\bf r}). 
\end{equation}
Consider any varied path ${\bf r} + \delta {\bf r}$ such that the virtual path and the true path coincide at the two end points, that is, $\delta {\bf r} = 0$ at the end points.
Taking  dot product of variation $\delta {\bf r}$ into Eq.~(\ref{rceq3}), and integrating the result along a path between the two end points, we have 
\begin{equation}\label{rceq4}
\int_{t_{1}}^{t_{2}} m\,{ d^{2}{\bf r} \over dt^{2}}\cdot \delta {\bf r} dt = \int_{t_{1}}^{t_{2}} - e^{-2(E-V({\bf r}))/m c^{2}} \nabla V({\bf r}) \cdot \delta {\bf r} dt. 
\end{equation}
The left hand side of Eq.~(\ref{rceq4}) becomes, after integration by parts,
\begin{eqnarray}\label{rceq5}
\int_{t_{1}}^{t_{2}} m\, {d {\bf v} \over dt}\cdot \delta {\bf r} dt 
 & = &  m\, {\bf v} \cdot \delta {\bf r}\mid _{t_{1}}^{t_{2}} - 
\int_{t_{1}}^{t_{2}}  m\, {\bf v} \cdot {d \over dt} (\delta {\bf r)} dt \nonumber \\
 & = & - \int_{t_{1}}^{t_{2}} \delta ( {1\over 2}\, m v^{2}  ) dt . 
\end{eqnarray}
The right hand side of Eq.~(\ref{rceq4}) is 
\begin{equation}\label{rceq6}
\int_{t_{1}}^{t_{2}} - e^{-2(E-V({\bf r}))/m c^{2}} \nabla V({\bf r}) \cdot \delta {\bf r} dt =
- \int_{t_{1}}^{t_{2}} \delta ( {1\over 2}\, m c^{2} \, e^{-2(E-V({\bf r}))/m c^{2}} ) dt. 
\end{equation}
From Eqs.~(\ref{rceq4}-\ref{rceq6}), we have
\begin{equation}\label{rceq7}
\int_{t_{1}}^{t_{2}} \delta ( {1\over 2}\, m v^{2}-{1\over 2}\, m c^{2} \, e^{-2(E-V({\bf r}))/m c^{2}}) dt = 0. 
\end{equation}
Since $E$  and $m\, c^{2}$ are  constants, we choose the Lagrangian  as
\begin{equation}\label{largrangian}
L =  {1\over 2}\, m v^{2}- {1\over 2}\, m c^{2} \,  e^{-2(E-V({\bf r}))/m c^{2}} + {1\over 2}\, m c^{2} \, e^{- 2 E/m c^{2}}. 
\end{equation}
By the very meaning of the principle of relativity, laws of quantum mechanics hold in all inertial frames, even though their mathematical forms are not Lorentz covariant. With a view to keeping up Heisenberg's uncertainty principle in the new relativistic quantum mechanics, the chosen Lagrangian is  different from the previous one \cite{yshuang4}. 
Substituting this Lagrangian into the Lagrange equations of motion
\begin{equation}\label{largrangianEq}
{d\over dt} {\partial L \over \partial \dot x_{j}} - {\partial L \over
\partial x_{j}} = 0, \,\,\, {\rm where}
\,\,\,\, j=1,2,3,
\end{equation}
we obtain 
\begin{equation}\label{rceq8}
m\, {d\, v_{j} \over dt}= - e^{-2(E-V({\bf r}))/m c^{2}} {\partial\, V({\bf r}) \over \partial x_{j}}. 
\end{equation}
The above equation is just the equation of  motion Eq.~(\ref{rceq}), provided that 
$ E = m\, c^{2}\,\, {\it ln}\, \gamma +  V({\bf r}).$ 
The actual path of a relativistic particle ought to depend on the initial energy $E$ of the particle, since the equation of motion Eq.~(\ref{rceq}) depends on the speed (equivalently, energy)  of the particle. Therefore, it is natural that the Lagrangian depends on the initial energy $E$. 

 From this Lagrangian, 
 the canonical momentum is
\begin{equation}\label{canMoment}
 p_{j}={\partial L \over \partial \dot x_{j}}= m v_{j},\,\,\, {\rm where} \,\, j=1,2,3.
\end{equation}
Then, the Hamiltonian is
\begin{eqnarray}\label{Haniltonian}
H & = & \sum_{j}{\partial L \over \partial \xder_{j}}\, \xder_{j} - L \nonumber \\
& = & {{\bf p}^{2} \over 2\, m } + {1\over 2}\, m c^{2} \, e^{-2 (E- V({\bf r}))/m c^{2}} - {1\over 2}\, m c^{2} \, e^{-2 E/m c^{2}},
\end{eqnarray}
where  ${\bf p}=  m {\bf v}$.
 Since $ E = m\, c^{2}\,\, {\it ln}\, \gamma +  V({\bf r})$, the Hamiltonian $H$ is
 equal to the quantity ${1\over 2}\,  m c^{2}\, (1- e^{-2E/m c^{2}} )$. Also, $H\approx {1\over 2}\, m v^{2} + V({\bf r})$, when $E<< m c^{2}$.

\section{\label{sec:6} A new theory of relativistic quantum mechanics }

   From the Hamiltonian Eq.~(\ref{Haniltonian}), with the standard procedure ---
 physical quantities in the Hamiltonian are replaced by their associated operators, ${\bf p} \rightarrow -i\hbar \nabla$ and ${\bf r} \rightarrow {\bf r}$, we have the new relativistic wave equation
\begin{eqnarray}\label{waveEq}
i\hbar {\partial \over \partial t} \Psi ({\bf r},t) & = & \hat{H}\, \Psi ({\bf r},t) \nonumber \\
 & = & -{\hbar^{2}\over 2m} {\bf \nabla }^{2}\, \Psi ({\bf r},t)   + {1\over 2}\, m c^{2} \,e^{-2E/m c^{2}} ( e^{ 2 V({\bf r})/m c^{2}} -1 )\, \Psi ({\bf r},t).
\end{eqnarray}

With the wave function $\Psi({\bf r},t)$ being normalized, $\Psi^{*}({\bf r},t)\, \Psi ({\bf
r},t)\, d^{3}{\bf r} $ is the probability of finding the particle in a volume
element $d^{3}{\bf r}$. It is immediately evident from Eq.~(\ref{waveEq})  that   the probability density
\begin{equation}\label{ProDensity}
 \rho = \Psi^{*}({\bf r},t)\, \Psi ({\bf r},t)
\end{equation}
and the probability current density
\begin{equation}\label{ProCurr}
{\bf J} = {\hbar \over 2mi}\, (\Psi^{*}\, {\bf \nabla} \Psi -
\Psi\, {\bf \nabla} \Psi^{*})
\end{equation}
satisfy the equation of continuity
\begin{equation}\label{ProEq}
 {\partial \over \partial t}\, \rho + {\bf \nabla} \cdot {\bf J} = 0.
\end{equation}
That is, the probability is conserved. According to this new theory of relativistic quantum mechanics, the probability density is positive definite and the statistical single-particle interpretation of quantum
theory is maintained.
In contrast,
the probability density  defined by the Klein-Gordon wave equation  in the current relativistic quantum
mechanics is not positive definite \cite{Bjorken, Greiner, Fuda}.

From the relativistic
wave equation Eq.~(\ref{waveEq}), the rate of the expectation value
of  position is
\begin{eqnarray}\label{derPos}
 {d \langle {\bf r} \rangle \over dt} & = & {d\over dt}\int \Psi^{*}({\bf
r},t)\,{\bf r}\, \Psi ({\bf r},t)\, d^{3}{\bf r} \nonumber \\
& = & \int \Psi^{*}({\bf r},t)\,{\bf r}\, {\partial \Psi ({\bf r},t)\over\partial t}\,
d^{3}{\bf r} + \int {\partial \Psi^{*} ({\bf r},t)\over\partial t}
\,{\bf r}\, \Psi({\bf r},t) \, d^{3}{\bf r} \nonumber \\ 
& = & \int \Psi^{*}\, (-i\hbar  \,\nabla \Psi )\,
d^{3}{\bf r} \, / m 
  =   \langle {\bf p} \rangle  \, / m .
\end{eqnarray}
This result $\langle {\bf p} \rangle = m\, d \, \langle {\bf r} \rangle / dt = m\,\langle {\bf v} \rangle$ is
consistent with the classical one ${\bf p}= m{\bf v}$.

For a free relativistic particle,  the rate of the expectation value
of  momentum is 
\begin{equation}\label{derMom}
{d\, \langle {\bf p} \rangle \over dt}  = \frac{1}{i\hbar} \langle\,  [\, \hat{\bf p}, \hat{H}\, ] \, \rangle
= 0,  
\end{equation}
because  the Hamiltonian of a free particle is $\hat{H} = \hat{\bf p}^2 / 2m $,  and $\hat{H}$ commutes with $\hat{\bf p}$. Thus, $\langle {\bf p} \rangle$ is a constant of motion for a free particle. Consequently, $\langle {\bf v} \rangle$ is a constant of motion for a free particle, since $\langle {\bf p} \rangle= m\,\langle {\bf v} \rangle$.
In contrast, the current relativistic quantum
mechanics predicts 
an anomalous phenomenon ---  a free particle trembles rapidly, even though its momentum is  constant --- the so-called {\it Zitterbewegung}  \cite{ Kalnay, Feshbach, Barut, Lock, Sidharth,Deriglazov}.

By the method of separation of variables, $\Psi ({\bf r},t) = \psi ({\bf r})f(t)$, from the relativistic wave equation  Eq.~(\ref{waveEq}), we have $f(t)=e^{-i\, {\cal E} \, t/\hbar}$, where ${\cal E} = {1\over 2}\,  m c^{2}\, (1- e^{-2E/m c^{2}} )$, and the time-independent relativistic wave equation 
\begin{equation}\label{TIdwaveEq}
  -{\hbar^{2} \over 2m} \, {\bf \nabla }^{2} \psi ({\bf r}) + {1\over 2}\, m c^{2} \, \, (\, e^{-2\,(E-V({\bf r}))/m c^{2}}-1)\, \, \psi ({\bf r})=0.
\end{equation}

  For a free particle in one-dimensional motion, from  Eq.~(\ref{TIdwaveEq}) with $V(x)=0$,  we have the plane wave function $\Psi_{k} (x, t) \sim  e^{i(k x-\omega t)}$,  where $k={mc\over \hbar }\, \sqrt{1 - e^{-2\, E/m c^{2}}  }= mv/ \hbar$, and $\omega = {\cal E} / \hbar = m c^{2}\, ( 1 - e^{-2\, E/m c^{2}} )/2\hbar =  m v^{2}/ 2 \hbar$ (since  $E = m\, c^{2}\,\, {\it ln}\, \gamma$). Thus, the phase velocity  of the plane wave is $ v_{p} = \omega / k  = v/2$.
Nonetheless, the group velocity of the wave packet is 
$ v_{g} = d\omega / dk =  v $. The group velocity of the wave packet is just equal to the classical velocity $v$ of the particle. 

The canonical energy ${\cal E}$ of a free particle is always non-negative, since $E = m\, c^{2}\,\, {\it ln}\, \gamma \geq 0$. This is consistent with the fact that energies of a free real particle are  not negative.  In contrast, according to the current relativistic quantum mechanics, there exist negative-energy states for a free particle \cite{Bjorken, Greiner}. The anomaly was resolved by the ad hoc interpretation of the negative-energy states  ---  "holes" in the occupied "sea of negative-energy states" are regarded as anti-particles. Yet this ad hoc interpretation disables the current relativistic quantum mechanics from maintaining the statistical {\it single-particle} interpretation of quantum theory. Also, this ad hoc interpretation can not be applied to bosons.
The  problems of the negative-energy states and  negative probability density cast a serious doubt on the current relativistic quantum mechanics, within the scope of statistical single-particle interpretation.

In addition, the canonical momentum of a particle of mass $m$ is $p=mv$.  Since $v<c$, the canonical momentum of the particle is less than $mc$. Thus, uncertainty of  momentum of the particle $\Delta p$  must be less than $mc$. Let $\Delta p = \alpha \, mc$, where $0 \le \alpha < 1$. From Heisenberg's uncertainty principle, $\Delta x \,\Delta p \gtrsim \hbar /2$, we have $\Delta x  \gtrsim  (1/4\pi\alpha) \lambda_{c} >  \lambda_{c}/4\pi$, where  $\lambda_{c} =  h/m c$ is the Compton wavelength. This implies that  the position of a particle of mass $m$ can  not be determined precisely with an uncertainty less than $\lambda_{c}/4\pi$. In contrast, there is no upper bound on uncertainty of the position of a particle. Thus, it is possible for a particle to have momentum with arbitrary precision such that $\Delta p \rightarrow 0$ ($\Delta x \rightarrow \infty$), whereas it is impossible to determine particle's position exactly, i.e., $\Delta x \rightarrow 0$. The position  and the momentum of a particle are not on an equal footing. The usual Lorentz transformation of space-time coordinates becomes useless as  particle's position can not be determined exactly. The usual Lorentz transformation of space-time coordinates is inconsistent with Heisenberg's uncertainty principle.  This  reinforces the novel perspective on relativistic transformation  --- relativistic transformation is the transformation of virtual four-displacement $\delta x^{\alpha}$ (equivalently, relativistic four-momentum) of a particle, instead of space-time four-coordinate $x^{\alpha}$. 

If $|E| << m c^{2}$ and $|V| << m c^{2}$, the new relativistic wave equation Eq.~(\ref{TIdwaveEq}) becomes the time-independent Schr{\"o}dinger wave equation in non-relativistic quantum mechanics,
\begin{equation}\label{TIdSchrEq}
  -{\hbar^{2} \over 2m} \, {\bf \nabla }^{2} \psi ({\bf r}) +  V({\bf r}) \, \psi ({\bf r})= E\,\psi ({\bf r}).
\end{equation}

\section{\label{sec:7} One-Dimensional Square Step  potential}

  We apply the new relativistic quantum mechanics to study standard problems in quantum mechanics. Consider a particle impinging on square step of  potential
\begin{equation}\label{PotStep}
V(x) = \left\{  \begin{array}{crr} 0   &, \,\, {\rm if}  &  x<0  \\
               V>0 & ,\,\, {\rm if}  &   x > 0 
                \end{array}
       \right.
  \end{equation}
\vskip 0.2cm
{\noindent
Case 1: $E>V$.} \\
    From  the time-independent relativistic wave equation Eq.~(\ref{TIdwaveEq}),  for $x<0$,
\begin{equation}\label{waveEqStep}
{d^{2}\psi (x) \over dx^{2} } + k^{2} \psi (x) = 0,
\end{equation}
where $k={mc\over \hbar }\, \sqrt{ 1-  e^{-2E/m c^{2}} }$, 
and for $x>0$,
\begin{equation}\label{waveEqStep2}
{d^{2}\psi (x) \over dx^{2} } + k'^{2} \psi (x) = 0, 
\end{equation}
where $k'={mc\over \hbar }\, \sqrt{ 1-  e^{-2(E-V)/m c^{2}}}$.
Then, the general solution  is
\begin{equation}\label{waveFunStep}
\psi (x) =\left\{  \begin{array}{crr} A\, e^{ikx} + B\, e^{-ikx} &, \,\, {\rm if}  &  x<0 \\
               C\, e^{ik'x}  &, \,\, {\rm if}  &  x>0 
                   \end{array}
\right.
\end{equation}
where $A$, $B$, and $C$ are constants.
Since the particle is incident from left in the region $x<0$,
the term $e^{-ik'x}$ is not included in the solution in the region
$x>0$. By matching  $\psi (x)$ and $d\psi (x)/dx$
across the step at $x=0$, we obtain 
$A+B=C$ and $k(A-B)=k'C$. Hence, 
\begin{equation}\label{reflect}
{B\over A} = {k-k'\over k+k'},
\end{equation}
and
\begin{equation}\label{trans}
{C\over A} = {2k\over k+k'}.
\end{equation}
From Eqs.~(\ref{reflect}) and (\ref{trans}),  
\begin{equation}\label{reflPlusTrans}
{|B|^{2}\over |A|^{2}} + {k' \over k} {|C|^{2}\over |A|^{2}} = 1.
\end{equation}
Also,  substituting the wave
function Eq.~(\ref{waveFunStep}) into the probability current density  Eq.~(\ref{ProCurr}), we have
\begin{equation}\label{currDensityStep}
J = \left\{ 
    \begin{array}{crr} (\hbar k / m)\, [\, |A|^{2}-|B|^{2}]  &, \,\, {\rm if}  &  x<0 \\
                (\hbar k' / m)\, |C|^{2} &, \,\, {\rm if}  &  x>0 
              \end{array}
\right. 
\end{equation}
Thus, by the conservation of probability, we have 
$(\hbar k / m)\, [\, |A|^{2}-|B|^{2}] =
               (\hbar k' / m)\, |C|^{2}$.
Consequently,
the reflection coefficient is
\begin{equation}\label{reflCoe}
R = {k\,|B|^{2}\over k\,|A|^{2}} = {(k-k')^{2}\over (k+k')^{2}}, 
\end{equation}
and the transmission coefficient is
\begin{equation}\label{transCoe}
T  = {k'\, |C|^{2}\over k\, |A|^{2}} = {4kk'\over (k+k')^{2}}. 
\end{equation}
Thus, $R+T=1$, that is, the
probability is conserved.

\vskip 0.2cm
{\noindent
Case 2: $E<V$.} \\
From Eq.~(\ref{TIdwaveEq}), for $x<0$,
\begin{equation}\label{waveEqStep3}
{d^{2}\psi (x) \over dx^{2} } + k^{2} \psi (x) = 0,
\end{equation}
where $k={mc\over \hbar }\, \sqrt{ 1-  e^{-2E/m c^{2}} }$,
and  for $x>0$,
\begin{equation}\label{waveEqStep4}
{d^{2}\psi (x) \over dx^{2} } - q^{2} \psi (x) = 0, 
\end{equation}
where $q ={mc\over \hbar }\, \sqrt{  e^{-2(E-V)/m c^{2}}-1 }$.
Then, the general solution is
\begin{equation}\label{waveStep2}
\psi (x) =\left\{ \begin{array}{crr}  A\, e^{ikx} + B\, e^{-ikx} &, \,\, {\rm if}  & x<0 \\
               D\, e^{-q x} &, \,\, {\rm if}  & x>0    
                  \end{array}
\right.
\end{equation}
where $A$, $B$, and $D$ are constants.
The term $e^{q x}$ is not included in the solution in the region
$x>0$, because this term diverges in the limit $x\rightarrow +\infty$.
By the boundary conditions that $\psi (x)$ and $d\psi (x)/dx$ are
continuous at $x=0$, we have
$A+B=D$ and $ik(A-B)=-q D$.
Then,  
\begin{equation}\label{reflect2}
{B\over A} =  {(1-iq /k)/ (1+iq /k)},
\end{equation}
and
\begin{equation}\label{trans2}
{D\over A} = {2/ (1+iq /k)}.
\end{equation}
From Eqs.~(\ref{waveStep2}) and (\ref{ProCurr}), the probability current density $J(x)=0$ in the region $x>0$.   The reflection coefficient is
$R={|B|^{2}/ |A|^{2}} = 1$. That is, particles are
totally reflected, if their kinetic energies are less than the potential energy. This prediction is consistent with the prediction of non-relativistic quantum mechanics.

However, according to the current relativistic quantum mechanics, suppose  $ V > 2\, mc^{2}$ and the  kinetic energy of the particle is in the region between 0 and $V - 2\,  mc^{2}$, then  the reflection coefficient $R>1$ and the transmission coefficient 
$T<0$  ---  the so-called Klein paradox  \cite{Klein, Wergeland,Dombey,Calogeracos}.  Moreover, the reflection coefficient becomes $R = 1$, if  the kinetic energy of the particle is higher than $V- 2\, mc^{2}$, but less than $V$. The  Klein paradox is counter to the meaning of probability and the conservation of probability.

\section{\label{sec:8} One-Dimensional Square barrier potential}

   Consider a particle incident on a square barrier
\begin{equation}\label{PotBarrier}
V(x) = \left\{  \begin{array}{crl} 0   &, \,\, {\rm if}  &  x<0  \\
               V>0 & ,\,\, {\rm if}  &   0 < x < L \\
                0   &, \,\, {\rm if}  &  x > L 
                \end{array}
       \right.
  \end{equation}
\vskip 0.2cm

{\noindent
Case 1: $E>V$.} \\
    From  the time-independent relativistic wave equation  Eq.~(\ref{TIdwaveEq}),  in the regions $x<0$ and $x>L$,
\begin{equation}\label{waveEqBarr}
{d^{2}\psi (x) \over dx^{2} } + k^{2} \psi (x) = 0,
\end{equation}
where $k={mc\over \hbar }\, \sqrt{ 1-  e^{-2E/m c^{2}} }$. 
In the region $0<x<L$,
\begin{equation}\label{waveEqBarr2}
{d^{2}\psi (x) \over dx^{2} } + k'^{2} \psi (x) = 0, 
\end{equation}
where $k'={mc\over \hbar }\, \sqrt{ 1-  e^{-2(E-V)/m c^{2}}}$.
Then, the general solution  is
\begin{equation}\label{waveFunBarr}
\psi (x) =\left\{  \begin{array}{crl} A\, e^{ikx} + B\, e^{-ikx} &, \,\, {\rm if}  &  x<0 \\
               C\, e^{ik'x} +  D\, e^{-ik'x} &, \,\, {\rm if}  &  0<x<L \\ 
                    F\, e^{ikx} &, \,\, {\rm if}  &  x > L 
                   \end{array}
\right.
\end{equation}
where $A$, $B$, $C$, $D$, and $F$ are constants.
 The continuity conditions of $\psi (x)$ and $d\psi (x)/dx$
at $x=0$ and $x=L$ yield 
$A+B=C+D$, $k(A-B)=k'(C-D)$, $C \,e^{ik'L} +  D\, e^{-ik'L}= F\, e^{ik L}$    and $ k'( C \,e^{ik'L} -  D\, e^{-ik'L})= k\, F\, e^{ik L}$. Solving $B$ and $F$, we have 
\begin{equation}\label{refAmpBarr}
{B\over A} = 2\, i \, (k'^2 - k^2)\, {\rm sin}\, (k' L) \left[ 4k\,k' \, {\rm cos}(k' L) - 2 \,i \, ( k^2 + k'^2)\, {\rm sin}(k' L)\right] ^{-1} ,
\end{equation}
and
\begin{equation}\label{transAmpBarr}
{F\over A} = 4k\,k' e^{- ik L}\left[ 4k\,k' \, {\rm cos}(k' L) - 2 \,i \, ( k^2 + k'^2)\, {\rm sin}(k' L)\right] ^{-1} .
\end{equation}
Consequently, the reflection coefficient is 
\begin{eqnarray}\label{refCoeBarr}
R  &=& {|B|^{2}\over  |A|^{2}} \nonumber \\ 
 &=&  \left( \frac{k^2 - k'^2 }{2\, k\, k'} \right)^2 \,  {\rm sin}^{2}(k' L)\, \left[ 1+  \left( \frac{k^2 - k'^2 }{2\, k\, k'} \right)^2 \,  {\rm sin}^{2}(k' L)\right] ^{-1}\, , 
\end{eqnarray}
and the transmission coefficient is 
\begin{eqnarray}\label{transCoeBarr}
T  &=& {|F|^{2}\over  |A|^{2}} \nonumber \\ 
 &=&  \left[ 1+  \left( \frac{k^2 - k'^2 }{2\, k\, k'} \right)^2 \,  {\rm sin}^{2}(k' L)\right] ^{-1}\, . 
\end{eqnarray}

\begin{figure}[ht]
\begin{center}
\includegraphics[width=0.45\textwidth, clip=]{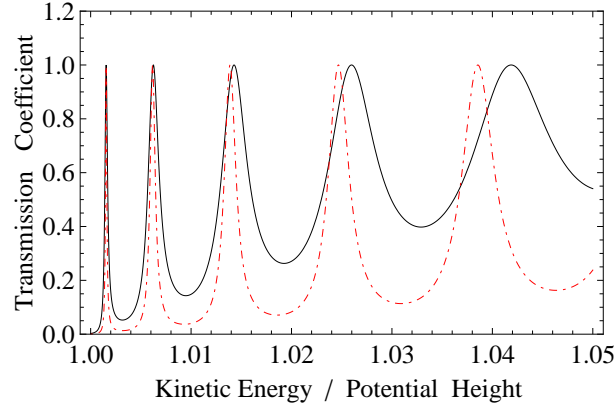}
\end{center}
\caption{\label{fig3}
Transmission coefficient of  particles impinging at square  barrier of height $V= 2\, m c^2$ and width $L = 40\,\, \hbar/ mc$. For a free particle of kinetic energy   $2\, m c^2$, its speed is  $\approx 0.99\,c$, as estimated by $ E = m c^2\, ln \gamma$. Solid line is the prediction by the new relativistic quantum mechanics. Dot-dashed line is the prediction by non-relativistic quantum mechanics.  }
\end{figure}

\begin{figure}[ht]
\begin{center}
\includegraphics[width=0.45\textwidth, clip=]{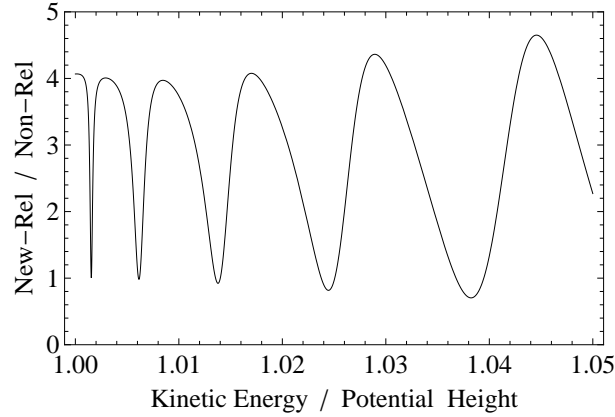}
\end{center}
\caption{\label{fig4}
Ratio of the transmission coefficients predicted by the new relativistic quantum mechanics  and non-relativistic quantum mechanics, for the case in Fig.~\ref{fig3} }. 
\end{figure}

 The formula of the transmission coefficient Eq.~(\ref{transCoeBarr})  is the same as that  by non-relativistic quantum mechanics \cite{Zettili}. Thus, the new relativistic quantum mechanics contains non-relativistic quantum mechanics as a low-speed limit, since  $k\approx \sqrt{2 m E}/ \hbar$ and $k' \approx \sqrt{2 m (E-V) }/ \hbar$ as $V<E<< mc^2$. As an example for the relativistic case, predictions of the new relativistic quantum mechanics  and non-relativistic quantum mechanics for the square barrier of height   $V= 2\, m c^2$ and width $L = 40\,\, \hbar/ mc$ are shown in Fig.~\ref{fig3}. Both predictions of transmission coefficient exhibit the resonance effect of tunneling as $k' L = n\, \pi$ (n is integer).    Ratio of the transmission coefficients predicted by these two theories  is shown in Fig.~\ref{fig4}. The transmission coefficient predicted by  the new relativistic quantum mechanics is averagely higher than  that by non-relativistic quantum mechanics. 

\vskip 0.2cm
{\noindent
Case 2: $E<V$.} \\
    From    Eq.~(\ref{TIdwaveEq}),  in the regions $x<0$ and $x>L$,
\begin{equation}\label{waveEqBarr3}
{d^{2}\psi (x) \over dx^{2} } + k^{2} \psi (x) = 0,
\end{equation}
where $k={mc\over \hbar }\, \sqrt{ 1-  e^{-2E/m c^{2}} }$. 
In the region $0<x<L$,
\begin{equation}\label{waveEqBarr4}
{d^{2}\psi (x) \over dx^{2} } - q^{2} \psi (x) = 0, 
\end{equation}
where $q ={mc\over \hbar }\, \sqrt{ e^{-2(E-V)/m c^{2}}- 1 }$.
Then, the general solution  is
\begin{equation}\label{waveFunBarr2}
\psi (x) =\left\{  \begin{array}{crl} A\, e^{ikx} + B\, e^{-ikx} &, \,\, {\rm if}  &  x<0 \\
               C\, e^{q x} +  D\, e^{-q x} &, \,\, {\rm if}  &  0<x<L \\ 
                    F\, e^{ikx} &, \,\, {\rm if}  &  x > L 
                   \end{array}
\right.
\end{equation}
where $A$, $B$, $C$, $D$, and $F$ are constants.
 The continuity conditions of $\psi (x)$ and $d\psi (x)/dx$
at $x=0$ and $x=L$ yield 
$A+B=C+D$, $i\, k(A-B)=q (C-D)$, $C \,e^{q L} +  D\, e^{-q L}= F\, e^{ik L}$    and $ q( C \,e^{q L} -  D\, e^{-q L})= i\ k\, F\, e^{ik L}$. Solving for $B$ and $F$, we have
\begin{equation}\label{refAmpBarr2}
\frac{B}{ A} = -i \frac{ k^2 + q^2}{ k\, q} \, {\rm sinh}(q L) \left[ 2\, {\rm cosh}(q L) + \,i \, \frac{ q^2 - k^2}{ k\, q}\, {\rm sinh}(q L)\right] ^{-1} ,
\end{equation}
and
\begin{equation}\label{transAmpBarr2}
\frac{F}{ A} = 2 e^{- ik L} \left[ 2\,{\rm cosh}(q L) + \,i \, \frac{ q^2 - k^2}{ k\, q}\, {\rm sinh}(q L)\right] ^{-1} .
\end{equation}
Consequently, the reflection coefficient $R$ and transmission coefficient $T$ are 
\begin{equation}\label{refCoeBarr2}
R   =  \left( \frac{ k^2 + q^2}{2\, k\, q}  \right)^2  {\rm sinh}^{2} (q L) \left[  {\rm cosh}^{2}(q L) +  \left( \frac{ q^2 - k^2}{2\, k\, q} \right)^{2}  {\rm sinh}^{2}(q L)\right] ^{-1}\, , 
\end{equation}
and
\begin{equation}\label{transCoeBarr2}
T  
 =  \left[ 1+  \left( \frac{k^2 + q^2 }{2\, k\, q} \right)^{2}  {\rm sinh}^{2}(q L)\right] ^{-1}\, . 
\end{equation}

\begin{figure}[ht]
\begin{center}
\includegraphics[width=0.45\textwidth, clip=]{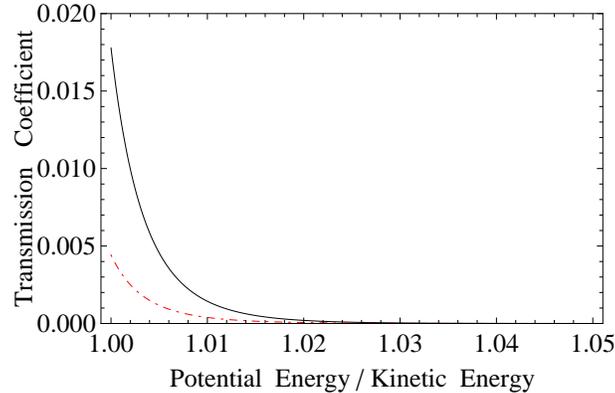}
\end{center}
\caption{\label{fig5}
The transmission coefficients versus the height of square barrier are predicted for the barrier of width $L = 15\,\, \hbar/ mc$ and  particle's kinetic energy $E= 2\, m c^2$.  Solid line is the prediction by the  new relativistic quantum mechanics. Dot-dashed line is the prediction by non-relativistic quantum mechanics.  }
\end{figure}

\begin{figure}[ht]
\begin{center}
\includegraphics[width=0.45\textwidth, clip=]{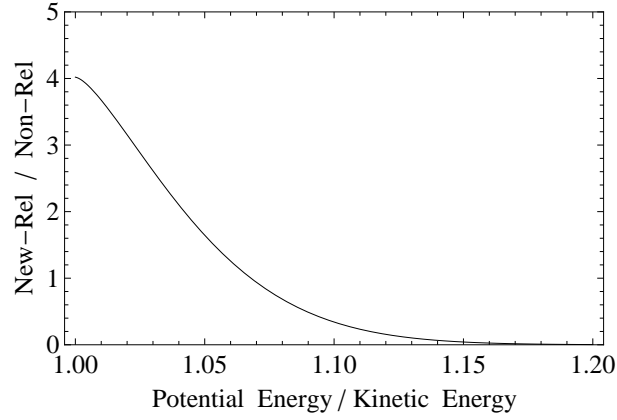}
\end{center}
\caption{\label{fig6}
Ratio of the transmission coefficient by  the new relativistic quantum mechanics  and non-relativistic quantum mechanics, for the case in Fig.~\ref{fig5}. The  transmission coefficient by this relativistic quantum mechanics is larger than that by non-relativistic quantum mechanics only when $V/E \lesssim 1.07$.  }
\end{figure}

 For particles of kinetic energy $E= 2\, m c^2$ and square barrier width  $L = 15\,\, \hbar/ mc$, predictions of the transmission coefficient versus the height of square barrier by the new relativistic quantum mechanics and non-relativistic quantum mechanics are shown in Fig.~\ref{fig5}. Both theories predict that the probability of tunneling decreases exponentially as the height of square barrier increases. In order to get larger transmission coefficient, the width of the barrier should be chosen as small as possible. Fig.~\ref{fig6} shows the ratio of the transmission coefficients by these two theories. The  transmission coefficient by the new relativistic quantum mechanics is not definitely larger than that by non-relativistic quantum mechanics, but only in a region of $V/E \lesssim 1.07$. This indicates that if one wants to experimentally detect the relativistic effect on the transmission coefficient, the height of barrier and particle's  kinetic energy  should be made as close as possible.


\begin{figure}[ht]
\begin{center}
\includegraphics[width=0.45\textwidth, clip=]{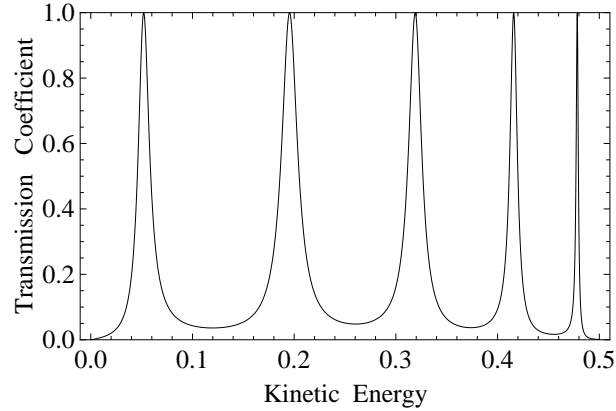}
\end{center}
\caption{\label{fig7}
Transmission coefficient versus kinetic energy is predicted by the current relativistic quantum mechanics for square barrier of height $2.5 \, m c^2$ and  width $L = 15\,\, \hbar/ mc$. The kinetic energy $ {\tilde E} - m c^2$ is in units of $m c^2$.   }
\end{figure}

 However, according to the Dirac wave equation in the current relativistic quantum mechanics, the reflection coefficient and the tranmission coefficient are  \cite{Dosch,Dombey,Calogeracos}
\begin{equation}\label{refCoeBarrDirac}
R =  \frac{ ( \, 1-{\tilde \kappa}^{2}\, )^2 \, {\rm sin}^{2}({\tilde q}\,  L) }{ 4 \, {\tilde \kappa}^{2} + ( \, 1-{\tilde \kappa}^{2}\, )^2 \, {\rm sin}^{2}({\tilde q}\,  L)}, 
\end{equation}
and
\begin{equation}\label{transCoeBarrDirac}
T   =  \frac{ 4 \, {\tilde \kappa}^{2}}{ 4 \, {\tilde \kappa}^{2} + ( \, 1-{\tilde \kappa}^{2}\, )^2 \, {\rm sin}^{2}({\tilde q}\,  L)}. 
\end{equation}
Here, $ {\tilde \kappa} = \sqrt{(V- {\tilde E} + mc^2)({\tilde E} + mc^2) }\, /  \sqrt{(V-{\tilde E} - mc^2)({\tilde E} - mc^2) }$ and $ {\tilde q} = -  \sqrt{(V-{\tilde E})^2 - m^2 c^4 }\, / \hbar c $. It should be noted that the energy of the particle ${\tilde E}$ includes the rest mass energy $mc^2$.  Eq.~(\ref{transCoeBarrDirac}) holds only for  $ V> 2\, mc^2 $ and  ${\tilde E}< V -m c^2$. In addition, from Eq.~(\ref{transCoeBarrDirac}), the transmission coefficient becomes unity, if ${\tilde q}\, L = n\, \pi$ ($n$ is integer). The tranmission coefficient is even not zero, as $V$ goes to infinity. These peculiar predictions are also called the Klein paradox. For  square barrier of width $L = 15\,\, \hbar/ mc$ and  height $V= 2.5\, m c^2$,
the transmission coefficient versus the kinetic energy $ {\tilde E} - m c^2$ predicted by  the current relativistic quantum mechanics is given in Fig.~\ref{fig7}. The kinetic energy of the particle is only  allowed in the region between 0 and 0.5 $mc^2$, from the definition of ${\tilde \kappa}$ and $V=2.5\, mc^2$.  There is a resonance effect in tunneling, even though the kinetic energy is in the non-relativistic regime and much less than the height of the barrier.  

\begin{figure}[ht]
\begin{center}
\includegraphics[width=0.45\textwidth, clip=]{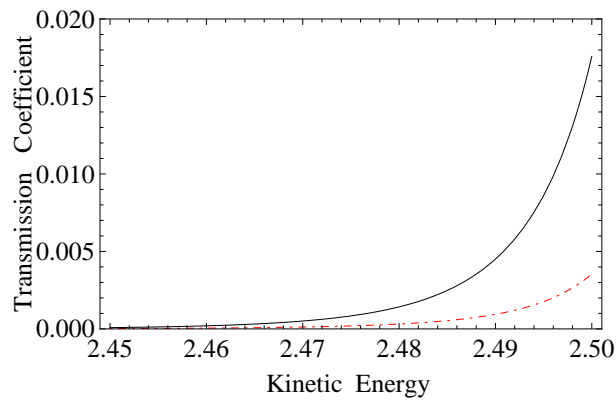}
\end{center}
\caption{\label{fig8}
Transmission coefficient for the same square barrier given in Fig.~\ref{fig7}. The kinetic energy is in units of $m c^2$. Solid line is the prediction by the new relativistic quantum mechanics. Dot-dashed line is the prediction by non-relativistic quantum mechanics.  }
\end{figure}

 For comparison, predictions by the new relativistic quantum mechanics  and non-relativistic quantum mechanics are given in Fig.~\ref{fig8}. These predictions are pronouncedly different from that by the current relativistic quantum mechanics. There is almost no tunneling for particles of kinetic energy less than $ 2.45\,m c^2$. The probability of tunneling raises exponentially as the kinetic energy increases  from  $ 2.45\,m c^2$ to $ 2.5\,m c^2$ (the height of the square barrier). Also, no resonance in tunneling is predicted by either the new relativistic quantum mechanics, or  non-relativistic quantum mechanics. 
The current relativistic quantum mechanics is contradictory to non-relativistic quantum mechanics even in the non-relativistic regime. Nevertheless, a true relativistic quantum mechanics must contain non-relativistic quantum mechanics as a low-speed limit. 

  Notwithstanding no experimental evidence confirming the Klein paradox, different reasons are given to rationalize the anomalies:  particle-antiparticle pair creation, hole theory, or virtual negative-energy incidence within the barrier \cite{Dombey, Calogeracos, Wergeland2, Alhaidari}. None of these reasons can resolve the Klein paradox beyond dispute. For instance,  by the pair creation,  the current relativistic quantum mechanics  should be replaced by quantum field theory. In that case, the current relativistic quantum mechanics is questionable. Therefore, the existence of the Klein paradox is dubious, and the  pair creation  by quantum field theory seems just an ad hoc excuse  for the artifical paradox. Furthermore, based on the pair creation, a new type of paradox, continual growth of anti-particles within the barrier, was pointed out \cite{Leo}. Oppositely, without appealing to quantum field theory, the Klein paradox can be avoided by proper-time formulation of relativistic quantum mechanics, within the single-particle interpretation \cite{Thaller,Fanchi,Horwitz0}. Even more, there may exist experimental evidence against the existence of the Klein paradox, as claimed recently \cite{Dragoman}. The Klein paradox has not been convincingly resolved for more than eighty years.

The Klein-Gordon wave equation and the Dirac wave equation in the current relativistic quantum mechanics are formulated by a systematic method based on the Lorentz covariance criterion --- forms of non-relativistic physical laws are modified into  Lorentz-covariant forms by an experiential guess methodology. Yet, a physical law need not be Lorentz-covariant to fulfill the principle of relativity. As well,  a 'physical law' need not fulfill the principle of relativity, even though its formula is made Lorentz-covariant. Since the principle of relativity can not be truly implemented by the Lorentz covariance criterion,  the Lorentz covariance criterion as currently applied in formulating the laws in relativity and relativistic quantum mechanics is  questionable. Thus, the validity of the Klein-Gordon wave equation and the Dirac wave equation should not be justified only by the Lorentz covariance criterion. Rather, owing to their inborn anomalies, the negative probability density, the negative-energy states and Zitterbewegung,  the Klein-Gordon wave equation and the Dirac wave equation are most likely wrong.

\section{\label{sec:9} One-Dimensional Infinite Square Well  }

 We apply the new relativistic quantum mechanics to study the other basic problems in quantum mechanics, the infinite and the finite square wells.   Consider a particle in an one-dimensional infinite square well
\begin{equation}\label{InfWell}
V(x) = \left\{  \begin{array}{crl} \infty   &, \,\, {\rm if}  &  x<-a  \\
               0 & ,\,\, {\rm if}  &   |x| < a \\
                \infty   &, \,\, {\rm if}  &  x > a 
                \end{array}
       \right.
  \end{equation}
In the region $|x|<a$, $V=0$ and the kinetic energy of the particle $E>0$. 
From the time-independent relativistic wave equation Eq.~(\ref{TIdwaveEq}), we have
\begin{equation}
\label{waveEqInfWell}
{d^{2}\psi (x) \over dx^{2} } + k^{2} \psi (x) = 0,
\end{equation}
where 
\begin{equation}\label{kInFinWell1}
k={m c \over \hbar }\sqrt{ 1- e^{-2E/m c^{2}}}.
\end{equation}
In the region $|x|>a$, we have $ \psi (x)=0$, since $V=\infty$.
Thus, the general solution is
\begin{equation}\label{waveFunInfWell}
\psi (x) = \left\{  \begin{array}{clc} A\, {\rm sin}(kx), \,\, {\rm or}\,\,  B\, {\rm cos}(kx)   &, \,\, {\rm if}  & |x|<a  \\
               0 & ,   &   {\rm otherwise}
                \end{array}
       \right.
  \end{equation}
where A and B are constant.
Since  $\psi (x)=0$ at the boundaries, we obtain $k = n \pi / L$, where $L=2a$ is the width of the well, and $n$ is an integer. Thus, we have the wave function for the bound states: $ {\rm sin}(n \pi x / L)$, where $n$ is an even integer, and $ {\rm cos}(n \pi x / L)$, where $n$ is an odd integer. From Eq.~(\ref{kInFinWell1}),  the 
energies of the bound states are
\begin{equation}
\label{EnerInfWell}
E_{n}=-{1\over 2}\, m c^{2}\,{\it ln}\, (1 - {n^{2} \pi^{2} \hbar^{2}  \over m^{2}c^{2}L^{2} } ),
 \end{equation}
where $n=1,2,3,...$.
 If $n^{2} \pi^{2} \hbar^{2}  / m^{2}c^{2}L^{2}  <<1 $, then
\begin{equation}
\label{EnerInfWell2}
E_{n} \approx {n^{2} \pi^{2} \hbar^{2}  \over 2 m L^{2} } \left[ 1+ \frac{1}{mc^{2}} \frac{n^{2} \pi^{2} \hbar^{2}}{2 m L^{2}} +\cdot \cdot \cdot \right]
 \end{equation}
 The eigen-energies of bound states  are $ {\bar E}_{n} = n^{2} \pi^{2} \hbar^{2} / 2 m L^{2} $, as predicted by non-relativistic quantum mechanics \cite{Zettili}. The eigen-energies $E_{n}$ of bound states by the new relativistic quantum mechanics are always larger than $ {\bar E}_{n}$.

\begin{figure}[ht]
\begin{center}
\includegraphics[width=0.45\textwidth, clip=]{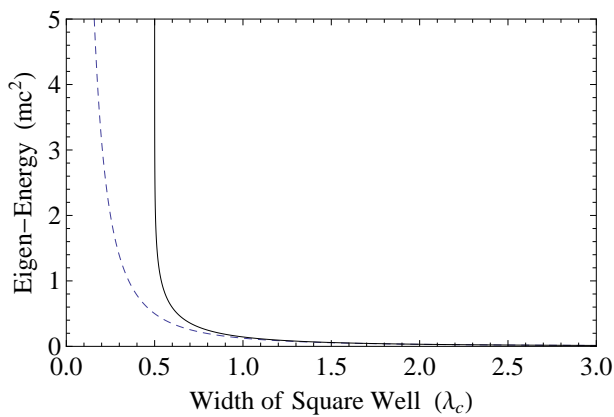}
\end{center}
\caption{\label{fig9}
The eigen-energy of ground state versus the width of infinite square well are predicted by
non-relativistic quantum mechanics (dashed line) and the  new relativistic quantum mechanics (solid line). The width of well is in units of Compton wavelength  $\lambda_{c} = h/m c$, and the eigen-energy is in units of $mc^2$.  }
\end{figure}
The eigen-energy of ground state versus the width of square well  predicted by the two theories is shown in Fig.~\ref{fig9}. Both theories show that the narrower the square well width is, the higher eigen-energy the ground state has. According to non-relativistic quantum mechanics, there are  bound states in an infinite square well, no matter how narrow the well width is. However, Fig.~\ref{fig9} shows  there is no bound state when the well width is less than half of the Compton wavelength  $\lambda_{c} =  h/m c$, in accordance with the new relativistic quantum mechanics. From Eq.~(\ref{EnerInfWell}), the eigen-energies $E_{n}$ are real, only if $n^{2} \pi^{2} \hbar^{2}/m^{2}c^{2}L^{2} \le 1$.  Therefore, if $L <  \pi \hbar/m c = \lambda_{c}/2$, then there is no bound state in the infinite square well. This is consistent with that, as pointed out in Sec.~\ref{sec:6}, the position of a particle of mass $m$ can not be determined precisely with an uncertainty less than $\lambda_{c}/4\pi$.

Though the problem of one-dimensional infinite square well is considered as the simplest, but essential, example in  non-relativistic quantum mechanics, this problem is not dealt with in textbooks of the current relativistic quantum mechanics \cite{Bjorken, Greiner}. It has even been commented that any consideration of the infinite square well is ruled out by the current relativistic quantum mechanics \cite{Coulter}. In spite of that comment,  eigen-energies of bound states are predicted by the Dirac wave equation with various assumptions, but the predictions are not the same \cite{Alberto, Alberto2,Alhaidari2}. It is fair  to say that the problem of one-dimensional infinite square well remains unsolved by the current relativistic quantum mechanics.

\section{\label{sec:10} One-Dimensional Finite Square Well }

  Now consider a particle in an one-dimensional square well with the potential
\begin{equation}\label{FinWell}
V(x) = \left\{  \begin{array}{crl} V   &, \,\, {\rm if}  &  x<-a  \\
               0 & ,\,\, {\rm if}  &   |x| < a \\
                V   &, \,\, {\rm if}  &  x > a 
                \end{array}
       \right.
  \end{equation}
We consider only bound states, thus  $0<E<V$.
From Eq.~(\ref{TIdwaveEq}), in the region $|x|<a$,
\begin{equation}\label{WaveEqFinWell1}
{d^{2}\psi (x) \over dx^{2} } + k^{2} \psi(x) = 0,  
\end{equation}
where
\begin{equation}\label{kFinWell1}
k={m c \over \hbar }\sqrt{ 1- e^{-2E/m c^{2}}},
\end{equation}
and in the region $|x|>a$,
\begin{equation}\label{WaveEqFinWell2} 
{d^{2}\psi (x) \over dx^{2} } - \kappa^{2} \psi(x) = 0,  
\end{equation}
where
\begin{equation}\label{kFinWell2}
 \kappa={m c \over \hbar } \sqrt{  e^{-2(E-V)/m c^{2}}-1}.
\end{equation} 
Also, the general solution is
\begin{equation}\label{waveFinWell}
\psi (x) =\left\{  \begin{array}{crl} A\, {\rm sin}(kx) , \, or\,\, B\, {\rm cos}(kx) &, \,\, {\rm if}  &  |x|<a \\
              C \,e^{-\kappa x} &, \,\, {\rm if}  &  x>a \\ 
                    D \,e^{+\kappa x} &, \,\, {\rm if}  &  x<-a 
                   \end{array}
\right.
\end{equation}
where $A$,  $B$, $C$ and $D$ are constants.
By matching  $\psi (x)$ and $d\psi(x)/dx$
at the boundaries, we have,  for even function solutions $\psi(x) =A\, {\rm cos}(kx)$, 
\begin{equation}\label{EvenFinWell}
k\,{\rm tan}(ka)=\kappa, 
\end{equation}
 and for odd function solutions $\psi(x) = B\, {\rm sin}(kx)$, 
\begin{equation}\label{OddFinWell}
k\,{\rm cot}(ka)=-\kappa. 
 \end{equation}

For a deep well,  $V >> mc^{2}$, from Eq.~(\ref{kFinWell2}) we have $\kappa >>1$. Thus, from Eqs.~(\ref{EvenFinWell}) and (\ref{OddFinWell}), $ka \approx n\pi/2$, where $n$ is an integer. Then, from Eq.~(\ref{kFinWell1}), we obtain $E_{n} \approx -{1\over 2}\, m c^{2}\, {\it ln}\, (1 - {n^{2} \pi^{2} \hbar^{2}  \over m^{2}c^{2}L^{2} } )$. This result is consistent with the eigen-energies of the infinite square well Eq.~(\ref{EnerInfWell}).

For a shallow well, $V<<mc^{2}$, we have $0<E<V << mc^{2}$. Then, from Eq.~(\ref{kFinWell1}), $k \approx \sqrt{2mE}/\hbar$, and from Eq.~(\ref{kFinWell2}), $\kappa \approx \sqrt{2m(V-E)}/\hbar$. Thus, in the non-relativistic regime, eigen-energies of bound states by the new relativistic quantum mechanics are approximately equal to those obtained by non-relativistic quantum mechanics \cite{Zettili}.

  From Eq.~(\ref{EvenFinWell}), we have
\begin{equation}\label{EvenFinWell2}
{\rm tan}(z) = e^{V/m c^{2}} \sqrt{(\frac{z_{0}}{z})^{2} -1}, 
\end{equation}
where $z\equiv k a$ and $z_{0}\equiv {m c a\over \hbar }\sqrt{ 1- e^{- 2V/m c^{2}}}$. From either Eq.~(\ref{EvenFinWell}), or Eq.~(\ref{EvenFinWell2}), we can solve the energies of even-parity bound states numerically.

\begin{figure}[ht]
\begin{center}
\includegraphics[width=0.45\textwidth, clip=]{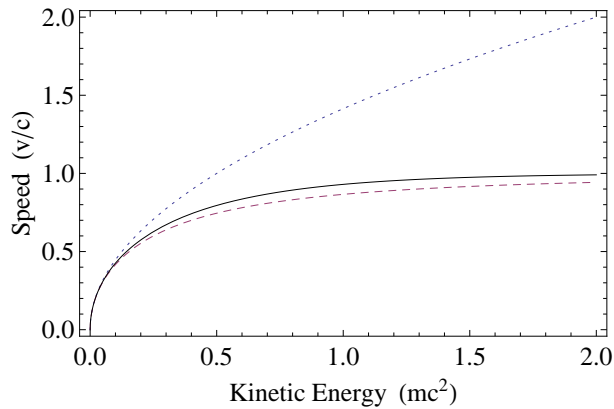}
\end{center}
\caption{\label{fig10}
The speed of a particle versus its kinetic energy: doted line is by $ E =  m v^2/2$ of non-relativistic theory, dashed line is by $ E = (\gamma -1) m c^2$ of the current relativity theory, and solid line is by $ E = m c^2\, ln \gamma$ of the  new relativity theory. The kinetic energy is in units of $mc^2$. The speed is in units of light speed.  }
\end{figure}

 As an example for a shallow well, let $V=0.3\, m c^{2}$ and $L=50\, \hbar / m c$.
The eigen-energies of bound states predicted respectively by non-relativistic quantum mechanics, Dirac theory \cite{Bjorken, Greiner,Coulter} and the new relativistic quantum mechanics are shown in Table~\ref{tab1}. It should be noted that the eigen-energies of bound states do not contain the rest mass energy. The results show that these three theories are consistent with each other in the non-relativistic regime $E \lesssim 0.1\, m c^2$. The relation between the speed and the kinetic energy of a particle is shown in Fig.~\ref{fig10}.  For  $E = 0.1\, m c^2$,  the speed of particle is:  $v  \approx 0.45\,c$ as estimated by $ E =  m v^2/2$ of non-relativistic theory; $v  \approx 0.42\,c$ as estimated by $ E = (\gamma -1) m c^2$ of the current relativity theory; or $v \approx 0.43\,c$ by $ E = m c^2\, ln \gamma$ of the new relativity theory  \cite{yshuang2}.  
\begin{table}[ht]
\begin{tabular}{|c||c|c|c|} \hline 
  &  \text{Non-Rel} &  \text{Dirac-Rel} &  \text{New-Rel} \\ \hline\hline
 1 & 0.0017846511 & 0.0017706025 & 0.0018135358 \\ \hline
 2 & 0.0160490225 & 0.0158215287 & 0.0165335149 \\ \hline
 3 & 0.0445045706 & 0.0433371366 & 0.0471525664 \\ \hline
 4 & 0.0869775503 & 0.0832348610 & 0.0962716984 \\ \hline
 5 & 0.1431072107 & 0.1340656791 & 0.1682948141 \\ \hline
 6 & 0.2120353124 & 0.1940810381 & 0.2674506448 \\ \hline
 7 & 0.2898509394 & 0.2608495741 & X  \\ \hline
\end{tabular}
\caption{\label{tab1}
The eigen-energies of even-parity bound states predicted respectively  by
non-relativistic quantum mechanics, Dirac theory and the  new relativistic quantum mechanics, for square well of height $V=0.3\, mc^2$ and width $L= 50\, \hbar /m c$. The eigen-energies are in units of $mc^2$. }
\end{table}

 For a deep well, let $V=2.3\, m c^{2}$ and $L=50\, \hbar / m c$ as an example.
The eigen-energies of bound states predicted respectively by these three theories are shown in Table~\ref{tab2}. 
\begin{table}[ht]
\begin{tabular}{|r||c|c|c|} \hline 
 &  \text{Non-Rel} &  \text{Dirac-Rel} &  \text{New-Rel} \\ \hline \hline
  1  & 0.001902293 & 0.359313042 & 0.001961917 \\  \hline
  2  & 0.017119946 & 0.445785446 & 0.017938337 \\ \hline
  3  & 0.047551517 & 0.537412903 & 0.051494736 \\ \hline
  4  & 0.093189369 & 0.633341306 & 0.106423759 \\ \hline
  5  & 0.154021609 & 0.732863972 & 0.190393795 \\ \hline
  6  & 0.230031505 & 0.835399176 & 0.319730324 \\ \hline
  7  & 0.321196599 & 0.940468560 & 0.536548760 \\ \hline
  8  & 0.427487394 & 1.047677969 & 1.014806876 \\ \hline
  9  & 0.548865387 & 1.156701163 & X \\ \hline
  10 & 0.685280096 & 1.267266244 & X \\ \hline
  11 & 0.836664369 & 1.379144391 & X \\ \hline
  12 & 1.002926666 & 1.492140340 & X \\ \hline
  13 & 1.183937482 & 1.606083954 & X \\ \hline
  14 & 1.379503294 & 1.720821912 & X \\ \hline
  15 & 1.589309931 & 1.836207756 & X \\ \hline
  16 & 1.812773705 & 1.952085926 & X \\ \hline
  17 & 2.048491237 & 2.068255025 & X \\ \hline
  18 & 2.288027319 & 2.184334004 & X \\ \hline
  19 & X & 2.297849281 & X \\ \hline
\end{tabular}
\caption{\label{tab2}
The eigen-energies of even-parity bound states predicted  respectively  by
non-relativistic quantum mechanics, Dirac theory and the  new relativistic quantum mechanics, for deep square well of height $V=2.3\, mc^2$ and width $L= 50\, \hbar /m c$. The eigen-energies are in units of $mc^2$.  For $E = mc^2$, the speed of particle is $v\approx 0.93\,c$ as estimated by $ E = m c^2\, ln \gamma$.}
\end{table}
The results show there is no bound state in the regime $ 0 < E <  0.3\, m c^{2}$, in accordance with Dirac theory \cite{Greiner,Coulter}.  In general, according to the current relativistic quantum mechanics, there is no bound state in the range  $ 0 < E <  V - 2\, m c^{2}$, as long as the height of the well $ V > 2\, m c^{2}$ --- the so-called Klein Paradox of deep square well \cite{Klein}.  The anomaly of the current relativistic quantum mechanics remains unresolved \cite{Kalnay,Wergeland,Dombey,Calogeracos}.  The current relativistic quantum mechanics does not contain non-relativistic quantum mechanics as a low-speed limit.

 In contrast, the results show that the new relativistic quantum mechanics is consistent with non-relativistic quantum mechanics in the non-relativistic regime. Yet, in the relativistic regime, the prediction by the new relativistic quantum mechanics is markedly different from that by  non-relativistic quantum mechanics. There are only a few bound states in the relativistic regime, as compared with the other two theories.  The reason is that the speed $v$ of the bound particle must be less than c, and the confining interaction on the particle is reduced by the factor $1-(v/c)^2$ \cite{YSHuang1,YSHuang2,YSHuang3}. These lead to an upper bound on energies of bound states, even for the infinite potential well.   

\section{\label{sec:11} Applications in nuclear physics}

  In nuclear $\beta$ decay, an electron is released from the nucleus. However, electrons are not  constituents of  the nucleus, according to the current theories and experimental data. That is, electrons can not be confined inside the nucleus by nuclear force. The current reason is as follows \cite{Yang,Das}: If an electron can be held inside a nucleus  whose radius is about 5 fm (this value is consistent with experimental data of the size of atomic  nuclei), then the maximum of wavelengths of the electron  inside the nucleus is $\lambda \approx 10$ fm. Thus, the momentum of the electron is $p = h/ \lambda \gtrsim 124$ MeV/c at least. By $E^{2}= p^{2}c^{2} + m^{2}c^{4}$ in special relativity and the rest mass energy of an electron $mc^{2}=0.511$ MeV, the energy of an electron inside the nucleus is at least about $E\approx pc \approx 124$ MeV. Yet, kinetic energies of electrons found in $\beta$-decay are in the range from KeV to a few MeV. Also, the nuclear binding energy per nucleon is about 8 MeV . Therefore, an electron can not be confined inside the nucleus by nuclear force, since the nuclear binding energy  is far less than 124 MeV. Nonetheless, this heuristic argument does not rule out the possibility that an electron can  be confined inside the nucleus in case the nuclear binding energy is sufficiently large.

 We  present an entirely different reason why an electron can not be confined inside the nucleus, based on the new relativistic quantum mechanics. As mentioned in Sec.~\ref{sec:9}, a particle of mass $m$ can not be confined in a region less than half of its Compton wavelength, even the confining potential is infinite. Therefore, an electron can not be confined inside the nucleus by whatever binding potential, since the Compton wavelength of an electron is about $\lambda_{c} \approx 2.4$ pm ($10^{-12}$ meter), whereas the radial size of the nucleus is $R\approx 1.2\, A^{1/3}$ fm ($10^{-15}$ meter) by experiments, where $A$ is nucleon number \cite{Yang,Das}.

  As a further implication, a nucleon  (proton, or neutron) can not be confined inside an infinite well of width less than 0.66 fm since the Compton wavelength of a nucleon is about 1.32 fm. Therefore, the diameter  of the nucleus is estimated as 0.66 fm at least, if the nucleus is simply regarded as an infinite square well confining nucleons. This estimation, though over-simplified, is consistent with  experimental data of the size of the nucleus.

 \section{\label{sec:12} Conclusion on the new relativistic quantum mechanics}
      
    The differential Lorentz transformation  is the transformation of virtual four-displacements $\delta x^{\alpha}$, or equivalently relativistic  four-momentum, of a particle.  Maxwell's equations of electrodynamics are rendered form-invariant through the transformation of electromagnetic fields in the k-space, rather than  in the space-time space.      The principle of relativity  means that  the same physical laws hold  in all inertial frames,  rather  than  their mathematical formulas  are Lorentz-covariant under the Lorentz transformation of  space-time coordinates. With the novel  perspective on relativistic transformation,  physical laws need not  satisfy the  Lorentz covariance criterion to  fulfill the principle of relativity. The  space and time concept underlying the new  relativistic transformation is Newtonian absolute space and absolute time. With the novel perspective on relativistic transformation and the very meaning of the principle of relativity,  quantum theory is  compatible with the principle of relativity.   

  A new theory of relativistic quantum mechanics is formulated based on  the  novel perspective on relativistic transformation. The underlying space and time concept of the new  relativistic quantum mechanics is Newtonian absolute space and absolute time. The relativistic quantum mechanics maintains the essentials of non-relativistic quantum mechanics, for example, the statistical single-particle interpretation and Heisenberg's uncertainty principle. Moreover, the new relativistic quantum mechanics contains  non-relativistic quantum mechanics as a low-speed limit, and it is free from the anomalies of the current relativistic quantum mechanics such as the negative probability density, the negative-energy states, the Klein paradox and  Zitterbewegung. In contrast, the current relativistic quantum mechanics does not contain  non-relativistic quantum mechanics as a low-speed limit.  Furthermore, a remarkable result is found as applying the new relativistic quantum mechanics to study the problems of square potential well --- a particle  can not be confined within an infinite square well of width less than half of the Compton wavelength of the particle. This finding implies that neither electrons, nor positrons, can be confined inside the nucleus.  Moreover, there is a lower bound on the size of the nucleus; the diameter of the nucleus is estimated as 0.66 fm at least .

  Einstein's notion of space-time is incompatible with the Newtonian notion of absolute space and absolute time. So far experimental tests of the Bell inequality have confirmed the predictions of quantum theory based on Newtonian absolute space and absolute time.   Supposing that experimental tests of  the Bell inequality  confirming quantum theory are indeed true, then experiments supporting Einstein's notion of space-time  become dubious. Besides experimental tests of the Bell inequality so far confirming quantum theory, the novel perspective on relativistic transformation and the very meaning of the principle of relativity necessitate a radical reappraisal of Einstein's notion of space-time and its experimental evidence.

\section{\label{sec:13}  Relativistic quantum statistics} 

Next based on this novel perspective, we formulate a new theory of relativistic quantum statistics, and then apply this theory to study thermal properties of a dilute gas. 

The postulates of non-relativistic quantum statistics are \cite{KHuang,Carter}:
\begin{enumerate}
\item  For an equilibrium system all micro-states are equally probable.
\item  The observed macro-state of a system is the one with the most micro-states.
\item  Pauli exclusion principle for identical particles.
\item  Boltzmann's postulate $S= k\,\, ln\, \Omega$.
\end{enumerate}
Based on the above postulates, the probability distribution functions for an equilibrium system  are derived as
\begin{equation}\label{disfun}
  f(\epsilon ) = \frac{1}{e^{(\epsilon - \mu) / k T} + a},
\end{equation}
where k is Boltzmann constant,  T is temperature, $\mu$ is chemical potential and $\epsilon$ is  energy of an eigen-state.  For Fermi-Dirac distribution, $a=1$;  for Bose-Einstein distribution, $a=-1$; for Maxwell-Boltzmann distribution, $a=0$. 

There are disputes on whether or not the three standard probability distribution functions are invariant, that is, the same formulas hold in all inertial frames. Based on the covariant proper-time approach,  alternative  relativistic probability distribution functions are derived to challenge  the invariance of the standard probability distribution functions \cite{Schieve}. However, that covariant proper-time approach is criticized from both theoretical and experimental aspects \cite{Debbasch}. In the experimental aspect, there are many experiments confirming the blackbody radiation that is formulated based on the standard Bose-Einstein distribution. This indicates that the Bose-Einstein distribution holds in all inertial frames. Thus, the covariant relativistic probability distributions are invalidated by experiments. It is very unlikely that the Bose-Einstein distribution is invariant, whereas the other two are not. All three standard probability distribution functions are invariant.

As extended to the relativistic realm, do the postulates and the three standard probability distribution functions  of non-relativistic quantum statistics hold in all inertial frames?
Postulates (1) and (2) are formulated in terms of the notion of statistics. Postulate (3) sets rules for identical particles. Postulate (4) provides a connection between statistics and thermal physics. These postulates are not expressed as mathematical formulas in terms of space and time parameters. That is, these postulates are formulated without any notion of space-time. Thus, whether or not these postulates fulfill the principle of relativity can not be decided by  their mathematical forms being (or not) covariant under Lorentz transformation of space-time coordinates --- the so-called Lorentz-covariant criterion. Furthermore, the probability distribution functions are derived just by maximizing the probability for the observed macro-state (or, maximizing the entropy for the equilibrium state in thermodynamics) based on the postulates \cite{KHuang,Carter}. The derivation has nothing to do with any notion of space-time. As mentioned previously, the same physical law can hold in all inertial frames (that is, the law fulfills the principle of relativity), even though its mathematical formula is not expressed in a Lorentz-covariant form. With this novel perspective,  we can assume that the postulates of non-relativistic quantum statistics fulfill the principle of relativity. That is, the postulates of non-relativistic quantum statistics are assumed to be the same in all inertial frames, though their forms are not manifestly Lorentz-covariant in the current perspective. The validity of this assumption can be justified by experimental tests on consequences derived from this assumption. Based on this novel perspective, we have a new theory  of relativistic quantum statistics --- all the postulates and all three probability distribution functions of {\it non-relativistic} quantum statistics are retained; yet the eigen-energy $\epsilon$ is evaluated by relativistic quantum mechanics, instead of non-relativistic quantum mechanics.

\section{\label{sec:14} Relativistic speed distribution of a dilute gas in a box}

 The first relativistic generalization of Maxwell speed distribution was presented by F. J{\"u}ttner in 1911 \cite{Juttner}.  J{\"u}ttner speed distribution has been accepted and applied in, for example, high-energy physics and astrophysics \cite{Chandrasekar,Groot}.  However, in recent years alternatives of relativistic speed  distribution were  proposed to challenge J{\"u}ttner speed distribution  \cite{Horwitz,Horwitz2,Schieve,Lehmann,Dunkel,Kaniadakis,Acosta}.  The correctness of these relativistic speed  distributions is still being actively investigated today \cite{Debbasch,Dunkel2,Kaniadakis2,Cubero,Ghodrat}. 

To derive speed distribution of a dilute gas, one has to evaluate eigen-energies and eigen-states of a particle in a box, and use them to derive the density of states. It should be emphasized that solving the problem of eigen-energies of a particle in a box is implausible by the current relativistic quantum mechanics  \cite{Bjorken,Greiner,Coulter}. However, by the new relativistic quantum mechanics, we are able to derive eigen-energies and eigen-states of a particle in a box, and then with the obtained results to derive the density of states. By the new theory of relativistic quantum statistics with the density of states obtained, we derive speed distribution of a dilute gas.

Consider a particle in a three-dimensional cubic box of width $L$. Solving the problem by Eq.~(\ref{TIdwaveEq}) of the new relativistic quantum mechanics, we have  the  wave functions of eigen-states
\begin{equation}\label{waveFunCubic}
\psi_{k_{x},k_{y},k_{z}} (x,y,z) = \left\{  \begin{array}{cl} A\, {\rm sin}(k_{x} x)\, {\rm sin}(k_{y} y)\, {\rm sin}(k_{z} z), &  {\rm inside\,\, the\,\, box}   \\
               0, &   {\rm outside\,\, the\,\, box} 
                \end{array}
       \right.
  \end{equation}
Here, $k_{x} = n_{x} \pi / L$, $k_{y} = n_{y} \pi / L$ and  $k_{z} = n_{z} \pi / L$, where $n_{x}$, $n_{y}$ and $n_{z}$ are positive integers. In addition,
\begin{equation}\label{kVectorCubic}
k_{x}^{2} + k_{y}^{2} +  k_{z}^{2} = \frac{ m^{2} c^{2}}{\hbar^{2}} \, (1- e^{-2E/m c^{2}} ).
  \end{equation}
The energy of the eigen-state $\psi_{k_{x},k_{y},k_{z}} (x,y,z)$  is 
\begin{equation}\label{eigenECubic}
 E(k_{x},k_{y},k_{z}) = - {m c^{2} \over 2} \,{\it ln} \left[\, 1- \frac{\hbar^{2} }{ m^{2} c^{2}}( k_{x}^{2} + k_{y}^{2} +  k_{z}^{2}) \, \right].
  \end{equation}

\begin{figure}[ht]
\begin{center}
\includegraphics[width=0.45\textwidth, clip=]{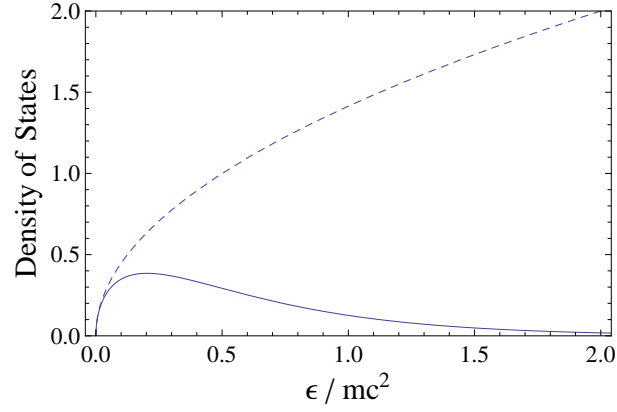}
\end{center}
\caption{\label{fig11} The non-relativistic density of states is shown in dashed line. This relativistic density of states is shown in solid line. Energy $\epsilon$ is in units of $mc^2$. The density of states is in units of $4\pi V m^3 c^3 / h^3$. }
\end{figure}

  By the well-known method  \cite{KHuang,Carter}, from  Eq.~(\ref{eigenECubic}) the density of states $g(\epsilon)$ is  
\begin{equation}\label{DOSrel}
g(\epsilon)  = \frac{4 \pi V m^{2} c}{ h^{3}} e^{-2\epsilon/ mc^{2}} \,  (1- e^{-2\epsilon/ mc^{2}})^{1/2},
  \end{equation} 
where $h$ is Planck constant, and $V=L^3$ is the volume of the box. 
In the low energy limit $\epsilon \ll  mc^{2}$, Eq.~(\ref{DOSrel}) reduces to the non-relativistic density of states of a particle in a box as derived by non-relativistic quantum mechanics \cite{Carter}
\begin{equation}\label{DOSonorel}
g_{c}(\epsilon)  = \frac{4 \sqrt{2} \pi V }{ h^{3}} m^{3/2}\, \epsilon^{1/2}  .
  \end{equation}
The non-relativistic density of states and the relativistic density of states are shown in Fig.~\ref{fig11}. The density of states is in units of $4\pi V m^3 c^3 / h^3$. The relativistic density of states and the non-relativistic density of states are almost indistinguishable when energies  $\epsilon \lesssim 0.05\, mc^2$.  The non-relativistic density of states is increasing monotonously with increasing energy. In contrast, the relativistic density of states is first increasing to the peak at an energy $\epsilon = ln(\sqrt{3/2})\, mc^2 \approx 0.20\, mc^2 $, and then is decreasing exponentially to zero. 

For a dilute gas, the Maxwell-Boltzmann distribution is re-expressed as 
\begin{equation}\label{MBdis}
f(\epsilon) = \frac{N}{Z} e^{- \epsilon / k\, T}, 
\end{equation}
where $N$ is the total number of particles of the gas, and $Z$ is the partition function
\begin{equation}\label{PartitionFun}
Z = \int_{0}^{\infty} g(\epsilon) \, e^{- \epsilon / k\, T} {\rm d} \epsilon.
\end{equation}
Substituting the density of states $g(\epsilon)$ in terms of $\gamma$, 
\begin{equation}\label{DOSrelGam}
g(\gamma)  = \frac{4 \pi V m^{3} c^{3}}{ h^{3}} \gamma^{-4} (\gamma^{2} -1)^{1/2},
  \end{equation}
into the above equation, we have the relativistic partition function 
\begin{equation}\label{PartitionFun2}
Z = \frac{\pi^{3/2} V m^{3} c^{3}}{h^{3}} \frac{\Gamma(\frac{1}{2\alpha} +1 )}{\Gamma(\frac{1}{2\alpha} + \frac{5}{2} )},
\end{equation}
where $\alpha \equiv k\, T / mc^{2}$ and  $\Gamma (x)$ is the gamma function \cite{Gradshteyn}.
The non-relativistic partition function derived by non-relativistic quantum mechanics is \cite{Carter}
\begin{equation}\label{PartitionNonRel}
Z_{c} = \left( \frac{2\pi m k T}{h^2} \right)^{3/2} V= \frac{\pi^{3/2} V m^{3} c^{3}}{h^{3}} (2 \alpha )^{3/2}.
\end{equation}
\begin{figure}[ht]
\begin{center}
\includegraphics[width=0.45\textwidth, clip=]{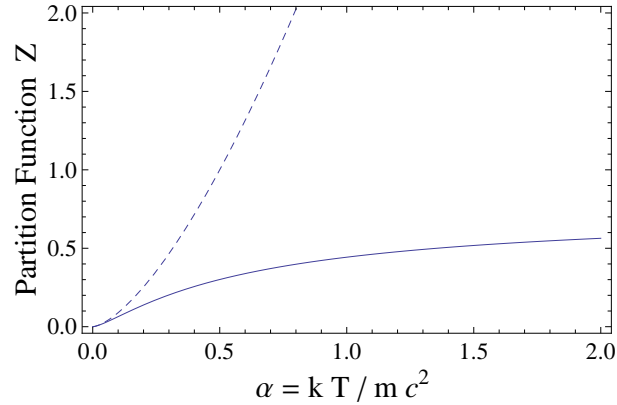}
\end{center}
\caption{\label{fig12}
 The non-relativistic partition function is shown in dashed line. This relativistic partition function is shown in solid line. The partition function is in units of $ V \pi^{3/2} m^3 c^3 / h^3$. }
\end{figure}
The non-relativistic partition function and this relativistic partition function Eq.~(\ref{PartitionFun2}) versus temperature are shown in Fig.~\ref{fig12}. Both partition functions are in units of $ V \pi^{3/2} m^3 c^3 / h^3 =  \pi^{3/2} V / \lambda_{c}^{3}$. This relativistic partition function reduces to the non-relativistic partition function when temperature  $\alpha \lesssim 0.01$.  As mentioned in Sec.~\ref{sec:9}, a particle can not be confined in an one-dimensional box of width less than half of the Compton wavelength. Thus, a particle can not be confined in a box of volume less than $(1/8) V_{c}$, where $V_{c}\equiv \lambda_{c}^{3}$, called Compton volume. One-eighth of the Compton volume is the least volume that can accommodate a particle. We consider $n_{Q}=\pi^{3/2} / V_{c}$ as quantum concentration. If the particle concentration  $n \equiv N/V$ is much less than the quantum concentration $n_{Q}$, then Maxwell-Boltzmann distribution can be applied in such case. 
Otherwise, Fermi-Dirac distribution should be used for Fermi gas, and Bose-Einstein distribution should be used for Bose gas. It should be noted that due to the density of states at high energy ($\epsilon \gtrsim 2\, mc^2$) being extremely low, one had better use Fermi-Dirac distribution for Fermi gas at extremely high temperature.

From Eqs.~(\ref{DOSrel}), ~(\ref{MBdis}) and ~(\ref{PartitionFun2}), the number of particles in the energy interval between $\epsilon$ and $\epsilon + {\rm d} \epsilon$ is
\begin{equation}\label{disDef}
\begin{array}{rcl}
N(\epsilon) \, {\rm d} \epsilon & = &  f(\epsilon) \,  g(\epsilon) \, {\rm d}\epsilon \\
 & = & \frac{4N}{\sqrt{\pi}} \frac{\Gamma(\frac{1}{2\alpha} + \frac{5}{2} )}{\Gamma(\frac{1}{2\alpha} +1 )} \,  e^{- 2 \epsilon /  mc^{2}} \, (1-e^{- 2 \epsilon /  mc^{2}} )^{1/2}\,  e^{- (\epsilon /  mc^{2})/ \alpha}  {\rm d} (\epsilon / mc^{2}).
\end{array}
  \end{equation}
In terms of $\gamma$, the distribution of particles is
\begin{equation}\label{disDefGama}
N(\gamma) \, {\rm d} \gamma  =  \frac{4N}{\sqrt{\pi}} \frac{\Gamma(\frac{1}{2\alpha} + \frac{5}{2} )}{\Gamma(\frac{1}{2\alpha} +1 )}  \gamma^{-4} (\gamma^{2}-1)^{1/2} e^{- {\it ln}\gamma /  \alpha} \, {\rm d} \gamma .
  \end{equation}
In terms of $\beta = v/c$, the relativistic speed distribution is
\begin{equation}\label{disDefBeta}
N(\beta) \, {\rm d} \beta  =  \frac{4N}{\sqrt{\pi}} \frac{\Gamma(\frac{1}{2\alpha} + \frac{5}{2} )}{\Gamma(\frac{1}{2\alpha} +1 )} \, \beta^{2}\, e^{- {\it ln}( 1 /\sqrt{1- \beta^{2}}) / \alpha}  \, {\rm d} \beta .
  \end{equation}
In the low temperature and the low speed limit, this relativistic speed distribution reduces to Maxwell speed distribution
\begin{equation}\label{disMaxBeta}
N_{M}(v) \, {\rm d} v  =  4 \pi N (\frac{m}{2 \pi k T})^{3/2}\,  v^{2}\, e^{- m v^{2} /2kT} \, {\rm d} v ,
  \end{equation}
because for $\alpha \ll 1$,
\begin{equation}\label{ApprGama}
\frac{\Gamma(\frac{1}{2\alpha} + \frac{5}{2} )}{\Gamma(\frac{1}{2\alpha} +1 )} \approx (\frac{1}{2\alpha})^{\frac{5}{2}-1}.
\end{equation}
Here, we apply the asymptotic representation of $\Gamma(x)$ for $|x| \gg 1$ \cite{Gradshteyn}, 
\begin{equation}\label{AsymGama}
 \Gamma(x) \approx x^{x-1/2}e^{-x} \sqrt{\pi}\, \{ 1 +\frac{1}{12 x}+\frac{1}{288 x^2} +...\}.
\end{equation}

\begin{figure}[ht]
\begin{center}
\includegraphics[width=0.45\textwidth, clip=]{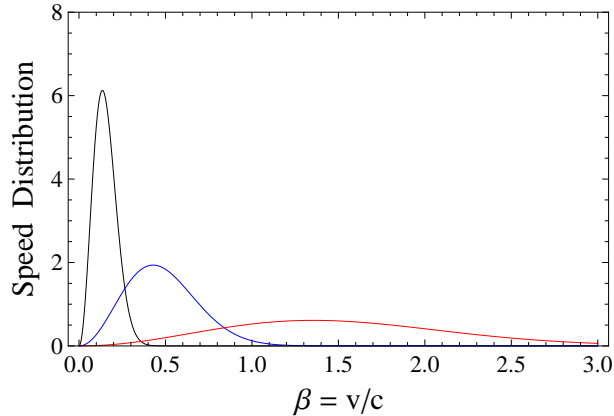}
\end{center}
\caption{\label{fig13} Maxwell speed distributions at temperatures $\alpha = 0.01$, $\alpha = 0.1$ and $\alpha = 1$ are shown, respectively, in the curves from left to right. The area under each Maxwell speed distribution is normalized to unity. }
\end{figure}

\begin{figure}[ht]
\begin{center}
\includegraphics[width=0.45\textwidth, clip=]{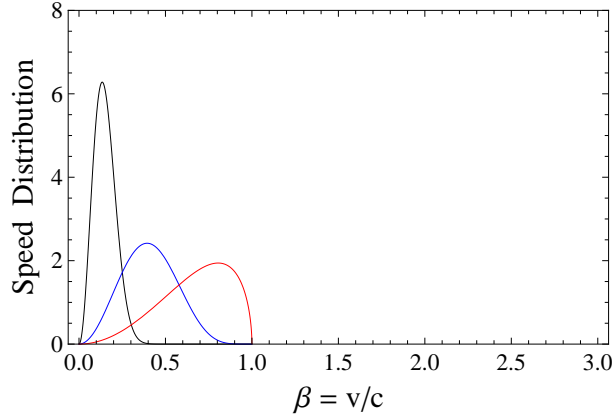}
\end{center}
\caption{\label{fig14} This relativistic speed distributions at temperatures $\alpha = 0.01$, $\alpha = 0.1$ and $\alpha = 1$ are shown, respectively, in the curves from left to right. The area under each relativistic speed distribution is normalized to unity. }
\end{figure}

Maxwell speed distributions of a dilute gas at various temperatures are shown in 
Fig.~\ref{fig13}. The curves from left to right are the speed distributions at temperatures $\alpha = 0.01$, $\alpha = 0.1$ and $\alpha = 1$, respectively. When $\alpha = 1$, there are more than half of the particles whose speeds exceed the speed of light. This relativistic speed distributions Eq.~(\ref{disDefBeta}) at the same various temperatures are shown in Fig.~\ref{fig14}. For all temperatures, there is no particle whose speed exceeds the speed of light. When $\alpha \lesssim 0.01$, this relativistic speed distribution and Maxwell speed distribution are almost indistinguishable. 
From Eq.~(\ref{disDefBeta}), the average speed of this relativistic speed distribution is
\begin{equation}\label{AveSpeed}
{\bar \beta} = \int_{0}^{1} \beta N(\beta) \, {\rm d} \beta / N = \frac{2}{\sqrt{\pi}}\frac{\Gamma(\frac{1}{2\alpha} + \frac{5}{2} )}{\Gamma(\frac{1}{2\alpha} + 3 )}. 
\end{equation}
For low temperature ($\alpha \lesssim 0.01$), ${\bar \beta} \approx (2 / \sqrt{\pi}) (1 / 2\alpha )^{5/2 -3} \approx \sqrt{8 k T / \pi m c^{2} } $; it is equal to the average speed of Maxwell speed distribution ${\bar v}  = \sqrt{8 k T / \pi m } $. For extremely high temperature,  ${\bar \beta}  \approx (2 / \sqrt{\pi}) (\Gamma(5/2) / \Gamma(3))  \approx 3/4 $. By similar calculation, the root-mean-square speed is $\beta_{rms} = \sqrt{3 \alpha / 1+ 5 \alpha }$. For low temperature, $\beta_{rms} \approx \sqrt{3 \alpha } \approx \sqrt{3 k T /  m c^{2} }$; thus $v_{rms}  \approx \sqrt{3 k T / m }$, the same as that of Maxwell speed distribution. For extremely high temperature, $\beta_{rms} \approx \sqrt{3/5}$.  The most probable speed is $\beta_{m}= \sqrt{2 \alpha / 1+ 2 \alpha }$. For low temperature, $\beta_{m} \approx \sqrt{2 \alpha } \approx \sqrt{2 k T / \ m c^{2} }$; thus $v_{m}  \approx \sqrt{2 k T / m }$, the same as that of Maxwell speed distribution.  For extremely high temperature, $\beta_{m} \approx 1$.

\section{\label{sec:15} New relativistic speed distribution versus J{\" u}ttner speed distribution}

\begin{figure}[ht]
\begin{center}
\includegraphics[width=0.45\textwidth, clip=]{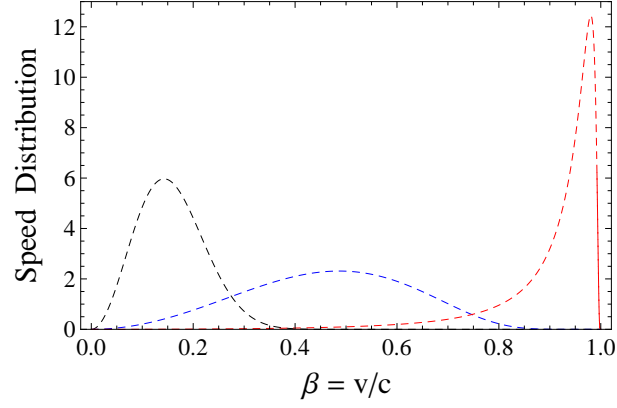}
\end{center}
\caption{\label{fig15} J{\" u}ttner speed distributions at temperatures $\alpha = 0.01$, $\alpha = 0.1$ and $\alpha = 1$ are shown, respectively, in the curves from left to right. The area under each J{\" u}ttner speed distribution is normalized to unity. }
\end{figure}

\begin{figure}[ht]
\begin{center}
\includegraphics[width=0.45\textwidth, clip=]{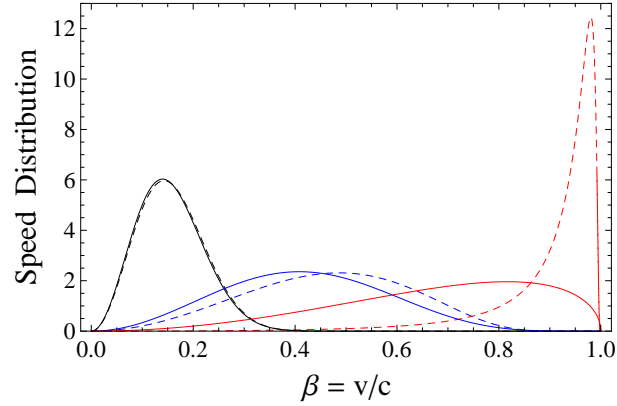}
\end{center}
\caption{\label{fig16} J{\" u}ttner speed distributions at temperatures $\alpha = 0.01$, $\alpha = 0.1$ and $\alpha = 1$ are shown, respectively, in the curves from left to right. The area under each J{\" u}ttner speed distribution is normalized to unity. For comparison, this relativistic speed distributions at the same temperatures are shown in solid line. }
\end{figure}

  The generally accepted relativistic distribution of a dilute gas is J{\" u}ttner  distribution \cite{Juttner,liboff}, in terms of $\gamma$,
\begin{equation}\label{JutdisGama}
N_{J}(\gamma) \, {\rm d} \gamma  =  \frac{N}{\alpha\, K_{2}(1/\alpha)}\,   \gamma \,(\gamma^{2}-1)^{1/2} \,e^{- \gamma /  \alpha} \, {\rm d} \gamma ,
  \end{equation}
where $K_{2}(x)$ is the modified Bessel function of order 2 \cite{Gradshteyn}. 
In terms of $\beta$, 
\begin{equation}\label{JutdisBeta}
N_{J}(\beta) \, {\rm d} \beta  =  \frac{N}{\alpha\, K_{2}(1/\alpha)}\,   \beta^{2}\, (1- \beta^{2})^{-5/2}\, e^{- \frac{1}{\alpha \sqrt{1- \beta^{2}}}}  {\rm d} \beta ,
  \end{equation}
J{\" u}ttner speed distributions of a dilute gas at various temperatures $\alpha = 0.01$, $\alpha = 0.1$ and $\alpha = 1$  are shown in 
Fig.~\ref{fig15}. Comparison between J{\" u}ttner speed distribution and this relativistic speed distribution is shown in Fig.~\ref{fig16}. The two relativistic speed distributions are remarkably different at high temperature $\alpha = 1$. According to J{\" u}ttner speed distribution, almost all the particles have speeds larger than $0.8\, c$. In contrast, according to this relativistic speed distribution, there are more than $67\, \%$ of particles whose speeds are less than $0.8\, c$. 

\section{\label{sec:16} Some thermal properties of a dilute gas }

    From this relativistic speed distribution, the total energy of the gas is
\begin{equation}\label{TotEngNew}
\begin{array}{rcl}
U  & = &  \int_{0}^{\infty}  \epsilon\, N(\epsilon) \, {\rm d} \epsilon =  \int_{1}^{\infty}  mc^{2}\, ln\gamma \, N(\gamma) \, {\rm d} \gamma  \\ & = & \frac{N\, mc^{2}}{Z} \int_{1}^{\infty}   ln\gamma \, g(\gamma) \, e^{ -ln\gamma / \alpha} {\rm d} \gamma . 
\end{array}
  \end{equation}
From the definition of partition function Eq.~(\ref{PartitionFun}), we have
\begin{equation}\label{DerPartition}
\left( \frac{\partial Z}{\partial T} \right) _{V} = \frac{k}{\alpha^2 mc^2} \int_{1}^{\infty}  ln\gamma \, g(\gamma) \, e^{ -ln\gamma / \alpha} {\rm d} \gamma . 
 \end{equation}
Comparing between the above two equations, the total energy is
\begin{equation}\label{TotEngNew2}
U =  N\, k\, T^2  \left( \frac{\partial\, ln Z}{\partial T} \right) _{V} . 
 \end{equation}
Furthermore, from Boltzmann's postulate $S= k\,\, ln\, \Omega$, with $\Omega = \Omega_{MB}$ subject to Maxwell-Boltzmann probability distribution, we have the entropy $S= \frac{U}{T} + Nk (lnZ -ln N +1)$. Also, we have the Helmholtz function  $ F=U-TS = -NkT (lnZ -ln N +1)$. 
From the above Helmholtz function, we have pressure $P=-(\partial F/ \partial V)_{T} = NkT (\partial \, lnZ/ \partial V)_{T}$. Then, using the relativistic partition function Eq.~(\ref{PartitionFun2}), we obtain  $PV =NkT$, the state equation of ideal gas. The state equation of ideal gas is valid for a dilute gas of relativistic particles at extremely high temperature.

From the total energy Eq.~(\ref{TotEngNew2}) and the relativistic partition function Eq.~(\ref{PartitionFun2}), we have
\begin{equation}\label{TotEngNew3}
U =  \frac{N\, mc^2}{2}  \left[ \psi(\frac{1}{2\alpha} + \frac{5}{2}) - \psi (\frac{1}{2\alpha} + 1) \right].
 \end{equation}
Here, $\psi(x)$ is the polygamma function $\psi(x) = \frac{{\rm d}}{{\rm d}\, x} ln\Gamma(x)$ \cite{Gradshteyn}. For low temperature,  we have $U \approx \frac{3}{2} NkT$, since $\psi(\frac{1}{2\alpha} + \frac{5}{2}) - \psi (\frac{1}{2\alpha} + 1) \approx 3\alpha$. For extremely high temperature $\alpha \gg 1$,  $U \approx \frac{1}{2} N\, mc^2  \left[ \psi( \frac{5}{2}) - \psi ( 1) \right]$. Then, we have $U \approx ( \frac{4}{3}- ln2)\, N\, mc^2 \approx 0.640186\,  N\, mc^2$.  The total energy of a gas enclosed in a finite region must remain finite, even if its temperature is raised extremely high.

\begin{figure}[ht]
\begin{center}
\includegraphics[width=0.45\textwidth, clip=]{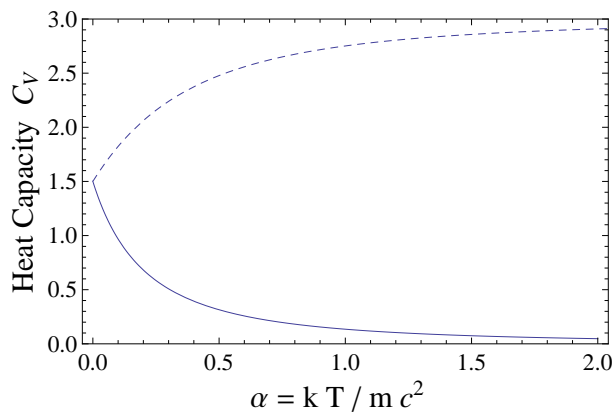}
\end{center}
\caption{\label{fig17} The heat capacity predicted by J{\" u}ttner speed distribution is in dashed line, and that by this relativistic speed distribution is in solid line.  The heat capacity is in units of Nk. }
\end{figure}

From Eq.~(\ref{TotEngNew3}), the heat capacity is
\begin{equation}\label{HeatCap}
C_{V}= \left( \frac{\partial U}{\partial T} \right) _{V} =  \frac{N\, k}{4\,\alpha^2}  \left[ \psi^{(1)}(\frac{1}{2\alpha} + 1) - \psi^{(1)}(\frac{1}{2\alpha} + \frac{5}{2})  \right].
 \end{equation}
Here, $\psi^{(n)}(x) = \frac{{\rm d^n}}{{\rm d}\, x^n} \psi(x)$ \cite{Gradshteyn}. For low temperature $\alpha \ll 1$,  we have $C_{V} \approx \frac{3}{2} Nk$, since $\psi^{(1)} (\frac{1}{2\alpha} + 1) - \psi^{(1)}(\frac{1}{2\alpha} + \frac{5}{2}) \approx 6\,\alpha^2$. For extremely high temperature $\alpha \gg 1$, $C_{V} \approx 0$. The heat capacity is decreasing to zero with increasing temperature. This is consistent with that the total energy of a system remains finite, even if its temperature is raised to infinite. In contrast, according to J{\" u}ttner speed distribution, $C_{V} \approx 3 Nk$ for extremely high temperature \cite{Lehmann,Acosta}. Comparison between the heat capacity by J{\" u}ttner speed distribution and that of this relativistic speed distribution is shown in Fig.~\ref{fig17}. For low temperature, both relativistic speed distributions predict $C_{V} \approx \frac{3}{2} Nk$, the same as that by Maxwell speed distribution. Yet, for extremely high temperature $\alpha \gg 1$, J{\" u}ttner speed distribution predicts that the heat capacity is increasing to $ 3 Nk$, whereas this relativistic speed distribution predicts that the heat capacity is decreasing to zero. It should be noted that predictions of the heat capacity by the current relativistic speed distributions are also pronouncedly different  \cite{Schieve,Lehmann}. So far, no experimental evidence distinguishing between these predictions is available \cite{Horwitz,Lehmann}. 

\section{\label{sec:17} Conclusion on the new relativistic quantum statistics}

All the current relativistic speed distributions are derived without using the current relativistic quantum mechanics, though Maxwell speed distribution can be derived in quantum statistics with non-relativistic quantum mechanics \cite{KHuang,Carter}. So far, relativistic eigen-energies of a particle in a box have not been solved by the current relativistic quantum mechanics. Thus, the density of states is untenable by the current relativistic quantum mechanics. This is probably the main reason why all the current relativistic speed distributions are not derived by using the current relativistic quantum mechanics.

 Theoretics of relativistic quantum statistics is far from complete and still being actively pursued \cite{Hamo,Cai}. 
Based on the novel perspective on relativistic transformation, a new theory of relativistic quantum statistics is formulated. By this new relativistic quantum statistics and the new relativistic quantum mechanics, a new relativistic speed distribution of a dilute gas is derived. This relativistic speed distribution reduces to Maxwell speed distribution in the low temperature region. This relativistic speed distribution is different from all the current relativistic speed distributions.  So far, no experimental evidence is available to determine which relativistic speed distribution is true.

\begin{acknowledgments}
We gratefully acknowledge Dr.  C.M.L.  Leonard and Dr.  B. Bruce for valuable comments
during the preparation of this paper.

\end{acknowledgments}

\end{document}